%% file: main.tex
\documentclass[letterpaper,twocolumn,10pt]{article}
\usepackage{usenix-2020-09}

\usepackage{tikz}
\usepackage{amsmath}

\usepackage{xcolor}
\def\BibTeX{{\rm B\kern-.05em{\sc i\kern-.025em b}\kern-.08em
T\kern-.1667em\lower.7ex\hbox{E}\kern-.125emX}}

\usepackage[mode=buildnew]{standalone}\usepackage{tikz}
\usetikzlibrary{patterns}
\usetikzlibrary{positioning,shapes.misc}
\usetikzlibrary{arrows,fit,calc,automata}

\usepackage[ruled,vlined]{algorithm2e} \usepackage{csquotes}
\usepackage{pgfplots}
\usepackage{pifont}
\usepackage{placeins}
\usepackage{multirow}

\usepackage{amssymb}
\usepackage{amsfonts}
\usepackage{amsfonts}
\usepackage{amsthm}
\usepackage{float}
\usepackage{scrextend}

\usepackage[binary-units=true]{siunitx}

\newcommand{\secache}{\textsc{ClepsydraCache}}

\usepackage{todonotes}
\begin{document}

\date{}

\title{\secache{} -- Preventing Cache Attacks with Time-Based Evictions}

\author{
{\rm
Jan
Philipp
Thoma}\\
Ruhr-University
Bochum
\and
{\rm
Christian
Niesler}\\
University
of
Duisburg
Essen
\and
{\rm
Dominic
Funke}\\
Ruhr-University
Bochum
\and
{\rm
Gregor
Leander}\\
Ruhr-University
Bochum
\and
{\rm
Pierre
Mayr}\\
Ruhr-University
Bochum
\and
{\rm
Nils
Pohl}\\
Ruhr-University
Bochum
\and
{\rm
Lucas
Davi}\\
University
of
Duisburg
Essen
\and
{\rm
Tim
Güneysu}\\
Ruhr-University
Bochum}
\maketitle
\definecolor{my_gray_light}{HTML}{e6e6e6}
\colorlet{my_red_light60}{red!60!white}
\colorlet{my_blue_light80}{blue!60!white}

\begin{abstract}
In
the
recent
past,
we
have
witnessed
the
shift
towards
attacks
on
the
microarchitectural
CPU
level.
In
particular,
cache
side-channels
play
a
predominant
role
as
they
allow
an
attacker
to
exfiltrate
secret
information
by
exploiting
the
CPU
microarchitecture.
These
subtle
attacks
exploit
the
architectural
visibility
of
conflicting
cache
addresses.
In
this
paper,
we
present
\secache{},
which
mitigates
state-of-the-art
cache
attacks
using
a
novel
combination
of
cache
decay
and
index
randomization.
Each
cache
entry
is
linked
with
a
Time-To-Live
(TTL)
value.
We
propose
a
new
dynamic
scheduling
mechanism
of
the
TTL
which
plays
a
fundamental
role
in
preventing
those
attacks
while
maintaining
performance.
\secache\  efficiently protects against the latest cache attacks such
as
\textsc{Prime+(Prune+)Probe}.
We
present
a
full
prototype
in
gem5
and
lay
out
a
proof-of-concept
hardware
design
of
the
TTL
mechanism,
which
demonstrates
the
feasibility
of
deploying
\secache\ in
real-world
systems.

\noindent\footnotesize\textit{A Clepsydra is an ancient time-measuring device worked by a flow of water.}
\end{abstract}
\section{Introduction}
\label{sec:intro}
\noindent
The
multi-layer
cache
hierarchy
is
a
fundamental
design
element
in
modern
microprocessors
that
bridges
the
performance
gap
between
the
main
memory
and
the
CPU.
By
keeping
frequently
accessed
data
in
close
proximity
to
the
CPU,
lengthy
pipeline
stalls
due
to
high
memory
latency
can
be
avoided.
Typically,
modern
desktop-grade
CPUs
implement
three
levels
of
cache
of
which
the
L1
and
L2
cache
are
unique
to
each
CPU
core,
while
the
last
level
cache
(LLC)
is
shared
among
all
cores.
The
size
of
cache
memory
is
usually
in
the
range
of
a
few
hundred
kilobytes
for
L1
caches
and
multiple
megabytes
for
last
level
caches.
Since
this
is
not
nearly
enough
memory
to
store
all
relevant
data,
caches
inevitably
need
to
evict
less
frequently
used
entries
to
make
space
for
new
data.

From
an
architectural
point
of
view,
caches
are
transparent
to
the
software,
i.e.,
a
process
does
neither
need
to
manage
data
stored
within
the
cache
nor
does
it
know
what
data
is
currently
cached.
However,
caches
store
data
based
on
temporal
locality.
That
is,
data
that
was
accessed
recently
is
cached
since
the
CPU
expects
it
to
be
accessed
again
soon.
Unfortunately,
the
story
of
software-transparent
caches
which
hide
the
latency
of
the
main
memory
was
disrupted
by
the
introduction
of
cache
timing
attacks~\cite{bernsteincache,
Osvik-2006-CacheAttacksandCo}.
Attackers
can
measure
the
execution
time
of
a
victim
program
or
even
a
single
memory
access
and
therefore
determine
whether
data
was
cached.
In~\cite{bernsteincache},
the
cache
side-channel
is
used
to
leak
an
AES
encryption
key
in
OpenSSL,
while~\cite{DBLP:conf/uss/YaromF14}
recovers
keys
from
a
victim
program
running
GnuPG.
Cache
side-channels
are
further
used
in
the
context
of
transient
attacks
like
Spectre~\cite{Kocher2018spectre}
and
Meltdown~\cite{Lipp2018meltdown}.
Even
software
running
in
a
trusted
execution
environment
like
Intel
SGX
can
be
attacked
using
cache
side-channels
as
demonstrated
in~\cite{Gotzfried-2017-CacheAttacksonInt,Brasser-2017-SoftwareGrandExpos}.
The
shared
nature
of
the
LLC
makes
it
a
particularly
worthwhile
target
for
attackers
since
the
timing
side-channel
can
be
exploited
beyond
process
boundaries
or
even
on
co-located
virtual
machines.
While
attacks
like
\textsc{Flush+Reload}~\cite{DBLP:conf/uss/YaromF14} and
\textsc{Flush+Flush}~\cite{10.1007/978-3-319-40667-1_14} exploit a special
instruction
that
allows
attackers
to
flush
specific
cache
lines,
the
\textsc{Prime+Probe}
attack
does
not
require
such
an
instruction
and
is
universally
applicable
across
ISAs~\cite{Osvik-2006-CacheAttacksandCo,DBLP:journals/joc/TromerOS10}.
Moreover,
in
contrast
to
flush-based
attacks,
the
latter
does
not
require
shared
memory
between
the
victim
and
the
attacker.
\relax

Since
the
very
design
goal
of
caches
is
to
accelerate
slow
accesses
to
the
main
memory,
the
timing
side-channel
cannot
easily
be
mitigated
without
losing
the
crucial
performance
benefit
that
caches
provide.
This
is
also
why
cache
side-channels
are
spread
over
a
huge
variety
of
CPUs
including
Intel,
AMD,
ARM
and
RISC-V
processors~\cite{Irazoqui-2016-CrossProcessorCach,gonzalez2019replicating}.
Since
the
vulnerability
is
rooted
deeply
in
the
hardware,
it
is
especially
difficult
to
mitigate
it
on
the
software
level~\cite{van2018malicious,
Doychev-2017-Rigorousanalysisof}.
Detection-based
mechanisms~\cite{DBLP:conf/osdi/ChenMXHPSZ14,DBLP:conf/micro/0020V14,
DBLP:conf/host/FangDYDV18,
DBLP:journals/corr/abs-1907-03651,
DBLP:conf/micro/YanST16}
suffer
from
false
positives
since
the
cache
access
patterns
vary
drastically
between
software.
Hence,
most
recent
proposals
feature
hardware
modifications
that
change
the
way
new
entries
are
cached.
The
architectural
solutions
can
be
divided
into
two
classes:
cache
partitioning~\cite{DBLP:conf/hpca/LiuGYMRHL16,
Wang-2016-SecDCP,DBLP:journals/iacr/Page05,
10.1145/1655008.1655019,sanchez2012scalable,
xie2009pipp,qureshi2006utility}
splits
the
cache
memory
into
disjunctive
security
domains,
thereby
preventing
information
leaks
beyond
security
domain
boundaries.
The
goal
of
partitioning
is
to
prevent
attackers
from
observing
evictions
originated
from
other
security
domains
than
their
own.
On
the
other
hand,
index
randomization
designs~\cite{werner2019scattercache,Tan2020,8574585,Wang2007,DBLP:conf/micro/WangL08,DBLP:journals/micro/LiuWML16}
do
not
reduce
the
amount
of
cache
conflicts
but
instead
randomize
the
mapping
of
addresses
to
cache
entries.
This
way,
the
cost
of
finding
eviction
sets
-
that
is,
a
set
of
addresses
that
map
to
the
same
cache
entries
as
the
victim
address
-
is
drastically
increased,
preventing
efficient
\textsc{Prime+Probe}
attacks.
However,
recent
work~\cite{purnal2021systematic}
demonstrates
that
index
randomization
alone
is
insufficient
for
security
and
requires
frequent
rekeying.
Attackers
can
still
construct
generalized
eviction
sets
that
have
a
high
probability
of
evicting
the
target
address.
Unfortunately,
frequent
rekeying
is
not
practical,
as
it
induces
high
performance
overhead
and
effectively
causes
a
complete
cache
flush.

\noindent \textbf{Contributions.}
With
\secache{},
we
introduce
a
novel
and
secure
index-based
randomization
scheme
which
effectively
protects
against
state-of-the-art
cache
attacks
while
at
the
same
time
preserving
the
performance
of
traditional
caches.
This
includes
\textsc{Prime+Prune+Probe}~\cite{purnal2021systematic}, which
is
one
of
the
most
recent
and
sophisticated
attack
techniques.
We
show
that
the
combination
of
cache
decay
and
index
randomization
effectively
protects
against
modern
cache
attacks.
Our
first
line
of
defense
is
a
randomized
address
to
cache
entry
mapping
(index
randomization).
Moreover,
we
introduce
a
hardware-integrated
time-to-live-based
eviction
strategy
(cache
decay)
that
drastically
reduces
the
amount
of
cache
conflicts
and
renders
reliable
observation
of
such
conflicts
infeasible.
This
prevents
the
attacker
from
learning
useful
information
when
observing
cache
misses
and
hence,
building
eviction
sets.
Our
security
analysis
shows
that
\secache{}
provides
strong
security
properties
against
state-of-the-art
attacks
including
the
recent
\textsc{Prime+Prune+Probe}~\cite{purnal2021systematic}
which
bypasses
pure
index-randomization
schemes.
Meanwhile,
our
design
maintains
the
flexibility
and
scalability
of
traditional
caches.
\secache\ thereby avoids additional complexity on
the
critical
path
of
cache
accesses,
such
as
the
indirections
used
in~\cite{mirageGuru}.
In
contrast
to
other
works~\cite{werner2019scattercache,Tan2020,8574585,Wang2007,DBLP:conf/micro/WangL08,DBLP:journals/micro/LiuWML16},
in
Section
\ref{subsec:hw_efficient}
we
provide
a
proof-of-concept
CMOS
design
for
the
key
time-to-live
feature
that
CPU
developers
can
build
on
for
minimal
overhead
implementation
of
\secache{}.
In
Section~\ref{sec:eval},
we
evaluate
\secache{}
using
gem5~\cite{DBLP:journals/corr/abs-2007-03152}
and
provide
a
detailed
performance
analysis
using
state-of-the-art
benchmarks
including
Parsec~\cite{bienia2008parsec}
and
SPEC
CPU
2017~\cite{BibEntry2021Mar}.
Our
results
indicate
a
small
performance
overhead
of
$1.38\%$.

\section{Background}
\label{sec:background}
\noindent
In
this
section
we
provide
background
information
on
cache
architectures
and
cache
side-channel
attacks.

\subsection{Caches}
\label{sec:background-caches}
\noindent
Caches
are
small
but
highly
efficient
temporary
storage
components
that
are
located
in
close
proximity
to
the
CPU.
They
bridge
the
gap
between
the
slow
main
memory
and
the
CPU.
Typically,
a
hierarchical
approach
is
taken
for
cache
memory,
i.e.,
each
CPU
core
is
equipped
with
a
small
L1
and
a
slightly
larger
L2
cache.
On
multicore
systems,
all
cores
typically
share
the
LLC.

\begin{figure}[htb]
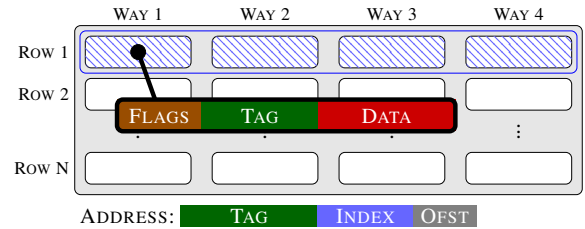

\centering
\includestandalone[width=.9\linewidth]{tikz/cache_addressing2}
\caption{\label{tikz:cache_addressing2}4-Way set associative cache with \textsc{N} sets (the first set is highlighted in blue). The flags section of an entry typically includes a valid and a dirty bit.}
\end{figure}

Since
caches
by
design
need
to
be
extremely
fast,
an
addressing
function
is
required
that
can
efficiently
determine
whether
the
requested
data
is
currently
cached,
where
it
is
located,
and
-
in
case
of
a
cache
miss
-
quickly
determine
which
entry
is
used
for
new
data.
Therefore,
most
architectures
use
set-associative
caches
which
feature
a
table-like
structure.
Set-associative
caches
are
divided
into
\textit{ways} and \textit{sets} which can be imagined
as
table
columns
and
rows,
respectively.
An
illustrative
layout
of
a
set
associative
cache
is
shown
in
Fig.~\ref{tikz:cache_addressing2}.
A
cache
\textit{entry}
is
a
cell
in
the
table
and
uniquely
addressed
by
the
\textit{set}
(in
the
non-randomized
setting
equivalent
to
the
row)
and
the
\textit{way}.
Ways
are
typically
managed
in
parallel
hardware
memory
blocks
(also
called
\textit{banks})
to
allow
concurrent
access
to
each
way.

When
a
program
performs
a
memory
operation,
the
virtual
address
is
first
translated
to
a
physical
address
by
the
memory
management
unit
(MMU).
The
lower
bits
of
the
virtual
address
indicate
the
page
offset
and
hence,
are
identical
to
the
physical
address.
The
upper
bits
of
the
virtual
address
hold
the
virtual
page
number
which
is
translated
to
a
physical
frame
number
in
the
physical
address.
Hence,
the
upper
bits
are
not
directly
controlled
by
user-level
software
running
on
the
processor.
For
cache
addressing,
the
physical
address
of
the
data
is
divided
into
a
\textit{tag},
an
\textit{index},
and
an
\textit{offset}.
If
an
entry
holds
$m$
bytes,
the
offset
part
of
the
address
is
$log_2(m)$
bits.
The
index
part
of
the
address
is
$log_2(N)$
bit,
where
$N$
is
the
amount
of
available
cache
sets.
The
remainder
of
the
address
is
used
as
a
tag
which
is
always
stored
alongside
the
data
in
cache
memory.
The
index
determines
the
set
(i.e.,
the
row)
in
which
the
data
is
stored.
Then,
all
entries
located
in
the
set
are
searched
for
the
tag
bits
of
the
incoming
address.
When
the
tag
matches
in
one
of
the
ways,
a
cache
hit
occurs,
and
the
memory
operation
is
performed
directly
in
the
cache.
If
the
tag
of
the
incoming
address
does
not
match
in
one
of
the
ways,
the
data
is
requested
from
another
memory
module
which
is
more
distant
to
the
CPU,
e.g.,
another
cache
or
the
main
memory.
When
the
request
is
served,
the
replacement
policy
selects
an
entry
within
the
set
determined
by
the
index
bits
and
stores
the
data
in
that
entry.
The
tag
of
the
address
is
again
stored
alongside
the
data.
If
the
entry
was
valid
and
a
write
was
performed
on
this
entry
prior
to
the
replacement,
the
data
needs
to
be
written
back
during
the
replacement
phase
since
the
modification
is
not
yet
committed
to
other
memory
modules
(writeback
dirty).
Each
entry
has
flags
indicating
whether
the
entry
is
modified
(\textit{dirty})
and
\textit{valid}.
Since
there
exist
diverse
variations
of
cache
memory,
including
different
store
and
writeback
policies,
we
refer
the
reader
to~\cite{hennessy2014computer}
for
a
detailed
description.

\subsection{Cache Side-Channels}
\label{sec:cache_side_channels}
\noindent
The
ability
to
distinguish
cache
hits
and
misses
based
on
the
latency
of
a
memory
instruction
can
be
leveraged
for
side-channel
attacks,
where
an
attacker
extracts
secret
information
from
other
processes
by
observing
the
cache
timing
behavior
and
therefore
can
draw
conclusions
about
accessed
data.

In
the
\textsc{Prime+Probe}
attack~\cite{Osvik-2006-CacheAttacksandCo,DBLP:journals/joc/TromerOS10},
an
attacker
fills
a
section
of
interest
in
the
cache
with
their
own
data
(\textit{prime}).
Next,
the
victim
process
is
executed
which
may
evict
some
of
the
attacker's
cache
entries
upon
memory
access.
In
the
final
step
(\textit{probe}),
the
attacker
measures
the
access
timing
for
re-loading
the
data
placed
during
the
prime
phase
to
recover
which
entries
were
evicted
by
the
victim.
Initially,
\textsc{Prime+Probe}
was
mostly
used
for
caches
closer
to
the
CPU,
since
the
cost
of
evicting
large
portions
of
the
LLC
is
typically
very
high.
However,
in~\cite{Liu-2015-Last-LevelCacheSid}
the
feasibility
of
\textsc{Prime+Probe}
attacks
on
LLCs
was
demonstrated
using
minimal
eviction
sets.
An
eviction
set
is
a
set
of
addresses
that
map
to
the
same
cache
set,
i.e.,
the
index
bits
of
the
address
are
equal.
An
eviction
set
is
minimal
if
the
amount
of
addresses
within
the
set
equals
the
number
of
cache
ways.
If
an
attacker
manages
to
create
such
an
eviction
set,
it
is
possible
to
efficiently
clear
selected
victim
data
from
all
cache
levels
including
the
LLC.
Due
to
the
translation
of
virtual
to
physical
addresses,
it
is
not
always
trivial
to
create
such
an
eviction
set
since
the
upper
bits
of
the
physical
address
depend
on
the
virtual
page
number
on
which
the
attacker
has
only
limited
influence.
To
overcome
this,
efficient
algorithms
for
obtaining
minimal
eviction
sets
have
recently
been
developed
~\cite{DBLP:conf/sp/VilaKM19,242066}.
\relax
A
variant
of
\textsc{Prime+Probe}
is
\textsc{Flush+Reload}~\cite{DBLP:conf/uss/YaromF14}.
While
this
attack
does
not
require
eviction
sets,
it
does
require
shared
memory
between
the
attacker
and
the
victim
as
well
as
a
\texttt{clflush}
instruction.
\relax

\relax

The
recent
\textsc{Prime+Prune+Probe}~\cite{purnal2021systematic}
targets
cache
architectures
that
use
index-randomization
to
prevent
the
construction
of
minimal
eviction
sets.
By
using
generalized
eviction
sets,
the
attack
can
reliably
evict
entries
from
the
cache
in
reasonable
time.
As
shown
later
in
this
paper,
the
attack
is
not
applicable
to
\secache{}.
\section{Problem Description and Related Work}
\label{sec:problem}
\noindent
The
fundamental
intrinsic
allowing
cache
attacks,
i.e.,
the
timing
difference
between
cache
hits
and
misses,
is
well
intended
and
in
fact,
the
very
reason
why
caches
exist.
Hence,
a
successful
mitigation
technique
must
hide
as
much
of
the
cache
internals
from
a
potential
attacker
without
forfeiting
the
performance
advantages.
Software
mitigations
for
cache
attacks
in
general
have
proven
to
be
costly
in
terms
of
computational
overhead
\cite{van2018malicious,
Doychev-2017-Rigorousanalysisof}.
Hence,
more
practical
approaches
usually
involve
hardware
changes,
sometimes
mediated
by
the
software.
Most
recent
mitigation
techniques
either
rely
on
cache
partitioning
or
index
randomization.

Cache
partitioning~\cite{DBLP:journals/iacr/Page05}
splits
a
cache
into
multiple
partitions,
which
are
assigned
to
different
security
domains.
Thus,
the
information
flow
between
different
partitions
is
constrained,
preventing
leakage
between
security
domains.
A
common
approach
is
to
divide
the
cache
among
its
cache
sets.
However,
this
approach
often
offers
only
one-way
protection~\cite{Wang-2016-SecDCP}
meaning
that
information
flow
from
a
confidential
to
a
public
partition
is
prohibited,
but
not
the
other
way
around
(public
to
confidential).
Generally,
a
distinction
is
made
between
static
and
dynamic
partitioning
schemes.
For
dynamically
partitioned
caches
like~\cite{sanchez2012scalable,
xie2009pipp,
qureshi2006utility},
timing
side-channel
attacks
are
possible
because
after
resizing
partitions,
entries
outside
of
the
shifted
boundaries
remain
valid.
Therefore,
the
cache
must
be
either
partitioned
statically
or
the
run-time
cache
partitioning
needs
to
be
designed
not
to
move
existing
entries
which
is
a
complicated
process.
The
disadvantage
of
static
partitioning
is
the
high
performance
overhead
resulting
from
the
limited
flexibility.
Since
cache
demands
vary
during
application
runtime,
the
allocated
partition
is
either
too
small
or
too
large
to
achieve
optimal
performance.
Furthermore,
the
amount
of
required
partitions
may
vary
within
short
time
periods
depending
on
the
machine's
workload.

A
different
approach
to
side-channel
secure
caches
is
based
on
randomizing
the
way
a
memory
address
is
mapped
to
a
particular
cache
set
and
line
(index
randomization).
For
instance,~\cite{DBLP:journals/micro/LiuWML16}
proposes
a
logical
direct-mapped
cache
with
extra
bits
for
the
indexing.
Those
extra
index
bits
significantly
increase
the
search
space
for
finding
cache
conflicts.
Recent
works
combine
the
randomization
approach
with
further
techniques
to
improve
the
security
against
ever-evolving
attacks.

\textsc{ScatterCache}~\cite{werner2019scattercache} uses a randomized cache mapping and
adds
software-assisted
domain
separation.
This
allows
duplicating
shared
addresses
in
the
cache
for
each
process
which
prevents
flush-based
attacks
like
\textsc{Flush+Reload}.
However,
as
shown
in~\cite{purnal2021systematic},
\textsc{ScatterCache} requires frequent re-randomization
to
avoid
construction
of
probabilistic
eviction
sets
to
protect
against
the
presented
\textsc{Prime+Prune+Probe}
attack.
All
designs
that
require
software
involvement,
including
partitioning
or
domain
separation
for
the
randomization
function
face
the
challenge
of
backwards
compatibility.
This
particularly
involves
two
features:
(1)
operating
system
adjustments
to
support
security
domains
and
(2)
support
for
legacy
software.
Therefore
by
default,
all
applications
belong
to
the
same
security
domain
if
the
software
is
not
aware
of
the
partitioning.

Another
approach,
dubbed
PhantomCache~\cite{Tan2020},
is
a
pure
architectural
solution
and
uses
a
localized
randomization
technique
to
bind
the
randomized
mapping
to
a
limited
number
of
cache
sets.
The
randomization
technique
used
in
PhantomCache
allows
an
address
to
be
mapped
to
multiple
locations
within
a
single
cache
bank.
Since
in
the
worst
case,
all
possible
entries
for
a
given
address
map
to
the
same
cache
bank,
parallel
lookup
is
not
possible.
Therefore,
the
lookup
latency
is
increased
over
traditional
randomization
schemes.

Recent
work~\cite{purnal2021systematic}
analyzed
the
security
properties
of
randomized
cache
designs
like
\textsc{ScatterCache}
and
PhantomCache
and
proposed
a
generic
attack
on
randomized
caches
dubbed~\textsc{Prime+Prune+Probe}.
The
attack
challenges
the
assumption
that
index-randomization
suffices
to
prevent
\textsc{Prime+Probe} attacks by constructing probabilistic eviction sets
with
relatively
small
sizes.
A
probabilistic
eviction
set
contains
addresses
that
collide
with
the
target
in
at
least
one
cache
way.
By
combining
multiple
such
addresses,
the
target
can
be
efficiently
evicted.
The
attack
is
split
into
a
profiling
and
an
attack
phase.
During
the
profiling
phase,
the
attacker
selects
a
set
of
addresses
$k$
and
accesses
them
repeatedly.
Due
to
the
randomization,
each
address
can
be
stored
at
an
independent
index
for
each
way.
This
leads
to
a
large
variety
of
combinations
where
the
$|k|$
addresses
can
be
stored
in
the
cache.
Eventually,
the
attacker
obtains
a
set
$k'\subseteq
k$
of
addresses
that
are
co-located
in
the
cache
without
evicting
each
other.
Then,
the
attacker
triggers
the
access
on
a
target
address
$x$
which
evicts
an
address
of
$k'$
with
\textit{catching
probability}
$p_c$.
If
such
an
eviction
is
observed,
the
evicted
address
from
$k'$
is
known
to
collide
with
$x$
in
at
least
one
cache
way
and
is
therefore
added
to
the
eviction
set
$G$.
This
phase
is
repeated
until
$G$
contains
a
sufficient
number
of
addresses.
For
an
attack
with
$90\%$
success
rate,
$G$
needs
to
have
between
36
(4-way
cache)
and
576
(16-way
cache)
addresses
for
schemes
where
each
way
has
an
independent
addressing
function,
see~\cite[Tab.
II]{purnal2021systematic}.
Once
$G$
is
established,
the
attacker
can
use
$G$
to
evict
$x$
similar
to
the
\textsc{Prime+Probe}
attack.

A
different
approach
for
side-channel
secure
caches
is
to
replicate
the
behavior
of
fully
associative
caches.
Since
those
designs
suffer
from
a
high
lookup
latency
for
large
caches,
they
are
usually
not
suited
for
large
LLCs.
The
recently
proposed
Mirage~\cite{mirageGuru}
mimics
a
fully
associative
cache
by
separating
the
tag
store
from
the
data
store.
By
over-provisioning
the
size
of
the
set-associative
tag-store,
conflicts
are
seldom.
Mirage
uses
a
bidirectional
pointer
mechanism
to
refer
from
the
tag
store
to
the
data
store
and
vice
versa.
However,
the
pointer
indirection
occurs
directly
on
the
critical
path
of
the
access.
Hence,
the
access
latency
of
Mirage
is
composed
of
$t_{rand}+t_{accessTag}+t_{accessData}$.
The
influence
of
$t_{rand}$
can
be
minimized
by
choosing
a
low-latency
randomization
function.
Since
the
tag-access
only
yields
the
location
for
the
data-access,
the
accesses
cannot
be
parallelized.
The
tag-store
is
managed
in
a
set-associative
structure
and
thus,
$t_{accessTag}$
is
similar
to
the
data
access
latency
in
a
traditional
cache
($t_{access}$).
If
we
assume
that
the
tag-
and
data-access
have
similar
latency,
the
overall
access
latency
of
Mirage
is
approximately
double
compared
to
traditional
caches:
$t_{rand}+t_{accessTag}+t_{accessData}
\approx
2\cdot
t_{access}$.
Besides
the
randomization
function,
\secache\ does
not
add
complexity
on
the
critical
access-path,
and
hence,
the
access
latency
in
hardware
is
closer
to
traditional
caches.
In
Section~\ref{subsec:hw_efficient}
we
propose
a
hardware
design
for
the
TTL
mechanism
of
\secache\ that
demonstrates
the
feasibility
of
efficient
implementations
of
such
designs
both
in
respect
to
performance
and
area.

\relax

\noindent
\textbf{Our Goal.}
To
summarize,
the
main
limitations
of
all
existing
proposals
are
that
they
either
do
not
prevent
all
attack
strategies
or
induce
high
performance
penalties.
To
prevent
cache
attacks,
the
goal
is
to
ensure
that
an
attacker
gains
no
exploitable
information
about
the
cache
content
while
preserving
the
efficiency
of
caches.
To
do
so,
we
develop
an
original
randomization
mechanism,
called
\secache\ that
reduces
information
leakage
far
beyond
currently
proposed
designs.
The
design
employs
a
low-latency
index-randomization
scheme
that
operates
on
the
entire
address
except
the
offset
bits.
The
cornerstone
of
\secache{}
is
an
eviction
strategy
based
on
timing
rather
than
contention.
This
strategy
is
also
known
as
\textit{cache decay} and has been considered for the purpose of
reducing
power
leakage
in~\cite{kaxiras2001cache}
and
for
access-trace-based
side-channels
on
AES
in~\cite{keramidas2008non}.
The
combination
of
randomization
and
cache
decay
is
unique
for
\secache{}
and
is
further
improved
by
a
dynamic
TTL
scaling
mechanism.
These
features
are
indispensable
for
the
security
against
state-of-the-art
attacks
including
\textsc{Prime+Prune+Probe}.
We
propose
the
first
hardware
design
of
the
cache
decay
feature
which
demonstrates
the
potential
of
low-overhead
implementations
of
\secache{}
by
CPU
developers.
\section{Threat Model}
\label{sec:threat}

\noindent
Our
threat
model
follows
previous
work
in
this
field~\cite{werner2019scattercache,
Tan2020,
purnal2021systematic}.
We
consider
a
black-box
attacker
in
an
ideal,
noise-free
scenario.
Hence,
the
attacker
is
able
to
perfectly
distinguish
between
a
cache
hit
and
a
cache
miss.
As
in
real-world
attacks,
the
attacker
has
no
insights
on
the
internal
state
of
the
cache
except
those
leaked
by
the
timing
of
memory
accesses.
Further,
the
attacker
is
able
to
access
an
arbitrary
number
of
addresses
and
measure
the
execution
time
of
a
victim
program,
potentially
revealing
the
cache
hit
and
miss
behavior
of
the
victim.
We
assume
the
index
randomization
function
to
be
pseudorandom,
i.e.,
the
attacker
cannot
predict
or
guess
the
cache
entry
to
which
an
address
is
mapped.
Attacks
on
the
hardware
level
are
out
of
scope
for
this
paper,
i.e.,
the
attacker
cannot
snoop
or
manipulate
the
memory,
memory
buses,
or
random
numbers.
While
we
consider
the
influence
of
the
chip
temperature
for
our
hardware
design,
attacks
tampering
the
physical
circuity,
or
the
environment
are
also
beyond
the
scope
of
this
paper.
Our
design
specifically
aims
to
prevent
conflict-based
attacks
like
\textsc{Prime+Probe}
and
deviates.
To
protect
against
flush-based
attacks
like
\textsc{Flush+Reload},
one
can
either
restrict
access
to
the
flushing
instruction
(\texttt{clflush}),
or
use
the
memory
duplication
method
presented
in~\cite{werner2019scattercache}.
We
investigate
in
detail
the
security
of
our
approach
in
Section~\ref{sec:security}.

\section{Concept}
\label{sec:concept}
\noindent
In
this
section,
we
provide
a
detailed
description
of
the
\secache{}
concept.
The
design
is
suited
for
all
levels
of
caches,
especially
including
shared
last-level
caches.

\subsection{\secache{} in a Nutshell}
\noindent
In
traditional
caches,
entries
can
either
be
evicted
using
a
special
instruction
defined
by
the
ISA
or
as
a
result
of
conflicting
addresses
that
are
mapped
to
the
same
entry.
All
current
cache
attacks
exploit
either
one
of
these
two
properties.
While
the
former
could
easily
be
addressed
by
not
offering
such
an
instruction
-
in
fact,
many
ISAs
do
not
feature
a
cache
line
flush
instruction
-
the
latter
is
more
of
a
fundamental
problem
in
current
cache
designs
and
cannot
easily
be
mitigated.

The
main
design
goal
of
\secache{}
is
to
drastically
reduce
the
overall
amount
of
cache
conflicts
and
to
remove
the
direct
linkage
between
cache
accesses
and
cache
evictions.
With
our
design,
we
achieve
this
goal
and
therefore
prevent
conflict-based
cache
attacks
including
\textsc{Prime+Probe}
and
\textsc{Prime+Prune+Probe}. \secache\ is compatible with
cache
duplication
of
shared
memory
as
proposed
in~\cite{werner2019scattercache}
to
defend
against
flush-based
attacks
like
\textsc{Flush+Reload}.

Instead
of
evicting
entries
when
a
conflicting
address
is
accessed,
in
\secache{}
each
entry
is
assigned
with
a
time-to-live
(TTL)
value
that
is
randomly
initialized.
The
TTL
is
steadily
reduced
and
when
expired,
the
entry
is
evicted
from
the
cache.
By
dynamically
adapting
the
global
rate
with
which
the
TTL
of
each
entry
decreases,
we
achieve
a
high
cache
utilization
with
minimal
conflicts.
Secondly,
\secache{} uses a highly randomized address to cache mapping,
and
in
doing
so,
efficiently
prevents
constructing
minimal
eviction
sets
without
observing
conflicts.

The
\secache{}
design
is
developed
with
large
caches
(i.e.,
typical
LLC
caches)
in
mind
and
therefore
must
respect
the
high
performance
requirement.
We
hold
on
to
the
traditional
set-associative
design
and
implement
a
variant
of
the
randomized
address-to-cache-mapping.
The
access-path
of
\secache\ is
therefore
equal
to
traditional
caches
with
the
addition
of
the
randomization
function.
Instead
of
checking
the
valid
bit
in
traditional
caches,
\secache\ needs
to
check
if
the
TTL
is
not
zero.
Similar
to
all
randomized
cache
architectures,
a
low-latency
randomization
function
is
key
to
a
low
access
time
of
the
cache.
The
TTL
mechanism
does
not
affect
the
critical
path.
As
studied
in
recent
years,
index-randomization
makes
finding
conflicting
addresses
much
more
challenging
for
an
attacker,
see~\cite{werner2019scattercache,
Tan2020}.
In
combination
with
the
time-based
eviction,
finding
eviction
sets
becomes
infeasible
as
we
discuss
in
detail
in
Section~\ref{sec:security}.

\subsection{Per-Entry Time-To-Live (TTL)}
\noindent
Each
entry
in
\secache{}
is
assigned
with
a
TTL
value
that
indicates
for
how
long
the
entry
remains
in
the
cache.
On
a
cache
miss,
the
accessed
data
is
loaded
from
the
memory
and
the
addressing
function
determines
the
target
entry
in
the
cache.
Simultaneously,
a
uniformly
random
TTL
in
between
a
lower
and
an
upper
bound
is
chosen
and
assigned
to
the
entry.
From
there
on,
the
TTL
is
steadily
reduced
with
a
reduction
rate
$\rm
R_{TTL}$.
When
the
TTL
for
an
entry
expires,
the
data
is
immediately
invalidated
and
-
if
required
-
written
back
to
the
main
memory.
If
a
cache
hit
occurs,
i.e.,
an
entry
with
$\text{TTL}>0$
is
accessed,
the
data
is
served
from
the
cache
and
the
TTL
is
reset
to
a
new
uniformly
random
value
between
the
lower
and
upper
bound.
This
is
to
make
sure
that
frequently
accessed
entries
stay
cached.
The
reduction
rate
$\rm
R_{TTL}$
globally
applies
to
all
entries
in
the
cache
and
is
scaled
dynamically
as
described
later
in
this
paper.

\subsection{Addressing and Replacement}
\label{subsec:addressing}
\begin{figure}[htb]
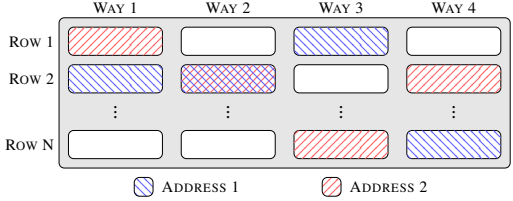

\centering
\includestandalone[width=.8\linewidth]{tikz/skewed_cache}
\caption{\label{tikz:skewed_cache}Address mapping of in \secache{} design. Each
address
maps
to
a
pseudorandom
cache
line
in
each
way.}
\end{figure}

\noindent
\secache{} implements a randomized address-to-cache-mapping.
In
traditional
caches,
the
index
bits
of
the
address
determine
the
line
in
which
the
data
is
placed,
such
that
the
entries
in
one
line
of
the
cache
form
a
set
(c.f.
Fig.~\ref{tikz:cache_addressing2}).
Therefore,
building
a
minimal
eviction
set
is
trivial
once
the
addressing
scheme
is
known
to
the
attacker.
In
\secache{},
we
use
a
low-latency
randomization
function
that
assigns
pseudorandom
cache
lines
in
each
way
to
a
given
address
as
shown
in
Fig.~\ref{tikz:skewed_cache}.
A
single
address
deterministically
maps
to
the
same
entry
in
each
respective
way
at
every
access.
For
an
address
$A$,
we
call
such
a
collection
of
indices
$A\rightarrow
\{i_1,...,i_w\}$ a \textit{dynamic set} that maps the address to an index in each
cache
way.
For
now,
we
assume
that
upon
access
each
dynamic
set
contains
at
least
one
invalid
entry
due
to
the
time-based
evictions.
One
of
the
invalid
entries
is
randomly
selected
to
store
the
data
without
causing
a
conflict.
We
consider
the
case
where
all
entries
in
a
dynamic
set
are
filled
(i.e.,
the
TTL
of
each
of
these
entries
is
greater
than
zero)
in
the
following
section.

Since
the
amount
of
cache
entries
is
limited
and
much
smaller
than
the
address
space,
addresses
with
conflicting
mappings
are
inevitable.
In
the
following
we
state
the
properties
of
an
ideal-keyed
addressing
function
that
is
indistinguishable
from
a
true
random
mapping
of
addresses
to
a
dynamic
set.
Such
a
function
$f_w:(\mathbb{F}_2^t,
\mathbb{F}_2^i)\rightarrow
(\mathbb{F}_2^{t'},
\mathbb{F}_2^i)$
is
instantiated
in
way
$w$
and
maps
a
$t$-bit
input-tag
and
$i$-bit
address-index
to
a
$t'$-bit
output-tag
and
an
$i$-bit
cache-index.
We
do
allow
$t
\neq
t'$
for
practicality
though
an
ideal
instantiation
has
$t
=
t'$.
For
an
address
$\rm
A=(tag||index||offset)$
in
a
$W$-way
cache
with
$N$
cache
lines
per
way,
$f$
should
yield
the
following
properties:

\begin{itemize}
\item \textbf{Invertibility:} Given a set $(tag, idx)$ in way $w$, it must be easy to
compute
$f^{-1}_w(tag,
idx)$.
This
is
required
for
writebacks
of
cache
entries
with
expired
TTL.
Invertibility
implies
injectivity,
i.e.,
there
must
not
be
a
set
of
tuples
$a=(tag,
idx)$
and
$b=(tag',
idx')$
with
$a
\neq
b$
such
that
$f_w(a)=f_w(b)$.
\item \textbf{Index-Pseudorandomness:} The probability that a set of tuples
$a=(tag,
idx)$
and
$b=(tag',
idx')$
with
$a\neq
b$
map
to
the
same
index
in
way
$w$
must
be
close
to
$1/N$
for
all
$w\in
W$.
It
must
be
difficult
to
construct
such
pairs.
Note
that
$N$
is
typically
small.
Hence,
this
is
not
a
genuine
hash
function.
\item \textbf{Independence:} For $w,w' \in W$ and $w \neq w'$, $f_w$ and $f_{w'}$ behave independently.
\end{itemize}

Note,
that
these
properties
do
not
require
a
collision-resistant
hash
function
or
a
cryptographically
secure
block
cipher.
The
reason
for
that
is
that
the
attacker
never
observes
the
actual
index
to
which
an
address
maps.
Instead,
the
only
observable
behavior
is
when
one
address
causes
an
eviction
of
another
address,
thus
indicating
that
both
addresses
collide
in
at
least
one
cache
way.

Hence,
we
leverage
a
round-reduced
block
cipher
with
strong
diffusion
characteristics.
The
addressing
function
in
\secache{}
is
interchangeable
and
related
work
proposes
several
different
techniques~\cite{werner2019scattercache,DBLP:journals/micro/LiuWML16,Tan2020}.
For
our
proof-of-concept
design,
we
choose
a
round-reduced
version
of
the
lightweight
block
cipher
PRINCE~\cite{DBLP:conf/asiacrypt/BorghoffCGKKKLNPRRTY12}
which
is
designed
to
provide
full
diffusion
-
that
is,
every
output
bit
depends
on
every
input
bit
-
after
two
rounds.
In
order
to
make
hardware
reverse
engineering
attacks
on
the
key
less
attractive,
the
secret
key
should
not
be
hardwired
for
an
implementation
but
instead
chosen
on
system
startup.

Using
a
block
cipher
for
the
address
randomization
comes
with
the
advantage
of
having
encrypted
tags
in
the
cache
and
hence
preventing
attacks
that
tamper
the
address
within
the
cache.
Traditional
block
ciphers
have
a
fixed
block
length
which
may
result
in
a
small
storage
overhead
for
the
tag
bits.
By
using
a
64
bit
block
cipher
like
PRINCE
the
ciphertext
has
a
length
of
64
bit
and
all
bits
of
the
ciphertext
are
required
for
the
decryption.
The
tag
bits
of
the
address
are
stored
just
like
in
traditional
caches
and
the
index
bits
of
the
encrypted
address
are
stored
implicitly
by
the
location
within
the
cache.
However,
in
traditional
caches
the
6
offset
bits
can
be
discarded.
When
a
64
bit
block
cipher
is
used,
the
offset
bits
can
be
zeroed
but
the
ciphertext
includes
6
bits
that
are
not
zero
and
need
to
be
stored
as
part
of
the
tag.
To
get
around
this
storage
overhead,
one
could
choose
a
format-preserving
encryption
(FPE)
scheme~\cite{bellare2009format,
bellare2010ffx,
schroeppel1998hasty}.
These
schemes
encrypt
an
$n$-bit
input
to
an
$n$-bit
output
and
are
hence
ideal
for
the
task.
Another,
likely
more
efficient
solution,
is
the
design
of
a
tailored
mapping
function
that
matches
the
security
requirements
and
the
ideal
block
size
exactly.

\relax

\subsection{Conflict Resolution}
\label{subsec:conflict_res}
\noindent
\secache{} dynamically controls the global speed with which the TTL of the entries
is
reduced
using
the
reduction
rate
$R_{TTL}$.
In
general,
a
high
$R_{TTL}$
means
that
entries
are
evicted
faster
while
a
low
$R_{TTL}$
yields
longer
lifetimes.
As
long
as
no
cache
conflicts
occur,
i.e.
a
dynamic
set
is
accessed
which
does
contain
empty
entries
\textit{or}
the
accessed
entry
is
already
cached
in
the
dynamic
set,
$R_{TTL}$
is
slowly
decreased
towards
the
minimum
value,
e.g.
$R'_{TTL}
=
R_{TTL}
-
1$.
When
a
conflict
occurs,
$R_{TTL}$
is
reset
to
a
higher
value,
e.g.
$R'_{TTL}
=
R_{TTL}\cdot
2$.
This
way
many
consecutive
conflicts
will
lead
to
fast
eviction
of
entries
(and
hence,
less
conflicts).

\begin{figure}[ht!]
\centering
\includegraphics[width=.8\linewidth]{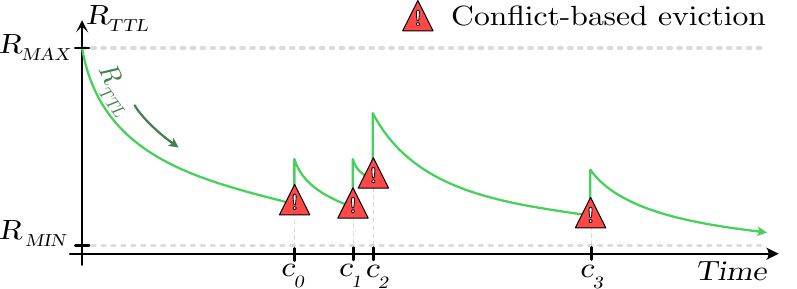}
\caption{\label{tikz:frequency}Evolution of the global TTL reduction rate that globally regulates the speed with which the TTL of each entry is reduced.}
\end{figure}

In
an
idealized
model,
one
would
adapt
$R_{TTL}$
such
that
every
dynamic
set
has
at
least
one
free
entry
at
all
times.
However,
this
would
require
expensive
monitoring
of
the
cache
utilization
in
each
dynamic
set.
Hence,
with
\secache{}
we
approximate
the
desired
behavior
by
dynamically
adapting
$\rm
R_{TTL}$
based
on
experienced
conflicts.
The
approach
is
comparable
to
TCP
congestion
control~\cite{allman1999tcp}.
Fig.~\ref{tikz:frequency}
illustrates
the
conceptual
evolution
of
the
TTL
reduction
rate
over
time.
We
start
by
setting
$R_{TTL}$
to
an
initial
value
and
over
time
decrease
it
slowly
towards
a
minimal
value,
i.e.,
the
lifetime
of
the
entries
increases.
If
a
dynamic
set
with
no
empty
entries
is
selected
on
a
cache
miss
(i.e.
a
conflict
occurs),
a
random
replacement
policy
is
used
to
replace
one
entry.
The
new
entry
is
assigned
with
a
uniformly
random
TTL
value.
Simultaneously,
$R_{TTL}$
is
increased
significantly,
shortening
the
lifetime
of
all
entries
which
increases
the
amount
of
empty
entries
and
hence,
reduces
the
probability
of
further
conflicts.
As
before,
$R_{TTL}$
is
slowly
reduced
towards
the
minimal
value
over
time.
This
results
in
a
shark-fin
shaped
evolution
of
the
reduction
rate
function
as
shown
in
Fig.~\ref{tikz:frequency}.
The
choice
of
the
initial
value
and
the
function
with
which
the
reduction
rate
develops
is
strongly
dependent
on
the
target
architecture
as
well
as
the
targeted
security
level
and
performance
and
hence
needs
to
be
optimized
individually.
We
describe
our
implementation
of
\secache{}
using
a
CPU
simulator
in
Section~\ref{sec:impl}.

\relax

\section{Security}
\label{sec:security}
\noindent
In
the
following
section,
we
evaluate
the
security
properties
of
\secache{}.
The
security
of
cache
architectures
using
index
randomization
to
thwart
side-channel
attacks
has
been
extensively
studied,
and
it
has
been
shown
that
finding
fully
congruent
eviction
sets
is
not
feasible
in
reasonable
time~\cite{werner2019scattercache,Tan2020,purnal2021systematic}.
However,
recent
work~\cite{purnal2021systematic,DBLP:conf/micro/BourgeatDYTEY20}
challenges
the
assumption
that
index
randomization
alone
prevents
cache
attacks
and
proposes
a
variant
of
\textsc{Prime+Probe}
called
\textsc{Prime+Prune+Probe}
which
relies
on
partially
congruent
eviction
sets.
\secache{}
implements
index
randomization
which
prevents
traditional
\textsc{Prime+Probe}
attacks
by
design
since
the
attacker
can
no
longer
construct
eviction
sets
trivially.
The
unique
combination
of
randomization
and
cache
decay
adds
a
second
layer
of
security,
the
benefits
of
which
we
will
outline
in
the
following.
Our
main
focus
lies
on
the
\textsc{Prime+Prune+Probe}
attack
since
this
is
the
most
relevant
attack
for
index-randomization-based
schemes.
We
consider
other
attack
vectors
in
Appendix
\ref{app:attacks}.

\subsection{Statistical Observations on TTL}
\label{sec:statistic_ttl}
We
start
the
security
analysis
of
\secache\ by
analyzing
how
an
attacker
may
distinguish
between
time-based
evictions
and
conflict-based
evictions.
The
ability
to
tell
those
types
of
evictions
apart
enables
the
attacker
to
gain
a
more
comprehensive
insight
to
the
cache
state
and
therefore
strengthen
its
capabilities.

In
traditional
caches,
a
conflict
is
detected
in
two
phases:
after
populating
the
cache
with
a
set
of
candidate
addresses,
the
attacker
accesses
a
target
address
which
may
evict
a
candidate
address
from
the
cache.
We
refer
to
this
phase
as
the
eviction
phase.
Subsequently
in
the
probe
phase,
the
attacker
searches
for
the
evicted
address
by
measuring
the
access
latency
to
all
candidate
addresses.
A
high
latency
in
that
phase
indicates
a
cache
miss
and
hence,
the
searched
conflict.
In
the
absence
of
noise,
this
works
for
an
arbitrarily
large
set
of
candidate
addresses.
In
\secache\ however,
the
introduction
of
random
time-based
evictions
introduces
a
level
of
uncertainty
for
the
attacker:
during
the
eviction
phase,
the
access
may
or
may
not
evict
an
address
from
the
cache.
Since
the
TTL
is
dynamically
balanced
to
maintain
empty
entries
in
the
cache,
in
most
cases,
no
entry
is
evicted
upon
access.
During
the
probe
phase,
the
attacker
then
accesses
a
set
of
prior
known-cached
addresses.
However,
when
they
observe
a
cache
miss,
there
is
no
way
to
distinguish
whether
the
missed
address
was
previously
evicted
based
on
timing
or
contention.
That
is,
since
the
attacker's
information
is
binary
(hit
or
miss)
and
does
not
reveal
whether
a
miss
was
based
on
an
expired
TTL
or
a
conflict.
This
breaks
the
direct
linkage
between
accesses
and
evictions
in
the
general
case.
However,
it
must
be
noted,
that
for
a
small
set
of
candidate
addresses,
the
time
between
the
eviction
and
the
observation
is
short
and
hence,
the
likelihood
of
a
time-based
eviction
is
low.
Therefore,
the
attacker
can
make
an
educated
guess
that
the
observed
cache
miss
was
caused
by
a
conflict
if
the
time
delta
between
the
initial
access
and
the
probe
step
is
small.
However,
we
show
in
Section
\ref{sec:ppp}
that
for
a
useful
attack,
the
attacker
must
probe
extensively
large
sets
of
candidate
addresses
which
make
time-based
evictions
much
more
likely
than
conflicts.

A
different
aspect
of
cache
conflicts
in
\secache\ is
that
a
sophisticated
attacker
might
be
able
to
monitor
cache
conflicts
during
the
eviction
phase
by
estimating
$R_{TTL}$.
Thereby,
they
aim
to
distinguish
cache
accesses
that
cause
a
cache
conflict
from
those
that
do
not.
To
achieve
this,
the
attacker
needs
to
access
a
set
of
addresses
and
periodically
probe
them
for
a
cache
miss.
Note
that
continuously
monitoring
a
single
address
is
not
feasible
since
a
hit
access
resets
the
TTL
of
the
respective
entry
to
a
new
random
value.
The
feasibility
of
this
heavily
depends
on
other
actions
taken
by
the
attacker,
the
precision
of
the
$R_{TTL}$
estimation,
and
the
overall
noise
level.
Even
if
the
attacker
can
observe
when
a
conflict
occurs,
they
cannot
know
during
the
probe
phase
\textit{which}
address
caused
the
conflict.
For
example,
if
the
attacker
knows
that
a
conflict
occurred
and
during
the
probe
phase
two
addresses
result
in
a
cache
miss,
they
cannot
distinguish
which
of
these
addresses
was
evicted
by
the
conflict
and
which
by
timing.
In
the
following,
we
assume
that
the
attacker
knows
if
an
access
caused
a
conflict
even
though
in
real
world
scenarios
this
may
not
be
feasible.

\subsection{Prime+Prune+Probe Attacks}
\label{sec:ppp}
\noindent
The
\textsc{Prime+Prune+Probe}~\cite{purnal2021systematic}
attack
described
in
Section~\ref{sec:problem}
builds
a
partially
congruent
eviction
set
$G=\cup_{i=1}^w
G_i$,
where
addresses
from
$G_i$
collide
with
the
target
address
$x$
in
way~$i$.
The
attack
consists
of
a
profiling
phase,
where
$G$
is
constructed,
and
an
attack
phase,
where
$G$
is
used
to
detect
cache
accesses
by
the
victim.
\secache{}
features
three
distinct
characteristics
that
prevent
efficient
\textsc{Prime+Prune+Probe}
attacks:
\begin{addmargin}[1em]{0em}
\ding{182} The priming set needs to occupy the entire set of the target address since
otherwise,
it
will
always
be
stored
in
the
empty
cache
entry.
This
increases
the
required
size
for
the
priming
set.\\
\ding{183} Time-based evictions add noise to the profiling and constrain the
maximum
time
between
two
accesses
to
each
address
of
the
priming
set.
Conflicts
during
\textit{prime}
and
\textit{prune}
further
reduce
this
timeframe
due
to
the
dynamic
scheduling
of
$R_{TTL}$.\\
\ding{184} The generalized eviction set must occupy all possible target entries since
otherwise,
the
target
will
always
be
stored
in
one
of
the
remaining
cache
entries.
Therefore,
the
eviction
set
needs
to
be
much
larger.
\end{addmargin}
In
the
following,
we
give
more
details
on
these
three
characteristics
and
compare
the
success
probability
of
\textsc{Prime+Prune+Probe}
using
\secache{}
to
pure
randomization
approaches
like
\textsc{ScatterCache}~\cite{werner2019scattercache}.
We
assume
that
no
noise
from
other
processes
is
present.
Finally,
we
give
a
conservative
complexity
estimation
of
profiling
attacks.
To
put
the
theoretical
security
results
into
context,
we
use
a
8\;MiB,
16-way
associative
cache
as
reference.
Similar
cache
configurations
can
be
found
in
recent
desktop-level
CPUs.
We
further
verified
our
theoretical
results
in
a
functional
Python
simulation.

\noindent\textbf{Catching Probability (\ding{182})}
During
the
\textit{prime}
and
\textit{prune}
phases,
the
attacker
repeatedly
accesses
a
set
of
addresses
$k$
and
removes
those
addresses
that
result
in
frequent
cache
misses,
resulting
in
a
set
$k'
\subseteq
k$
of
simultaneously
known-cached
addresses.
The
catching
probability
$p_c$
describes
the
probability
that
after
filling
the
cache
with
$|k'|$
addresses,
the
access
to
the
target
address
$x$
evicts
an
address
from
$k'$.

\secache{} is different from most other cache architectures in that the
cache
is
designed
to
never
be
completely
filled.
Though
an
attacker
can
deliberately
fill
many
cache
entries
by
accessing
a
huge
amount
of
addresses
in
a
very
short
time
frame,
the
global
TTL
scheduling
mechanism
is
designed
to
regulate
the
utilization
and
quickly
invalidate
entries
in
such
a
scenario.
The
TTL
mechanism
is
designed
to
balance
the
utilization
in
a
way
that
on
average,
each
dynamic
set
contains
at
least
one
free
entry
and
therefore,
the
amount
of
conflicts
is
minimized.
Importantly,
on
a
cache
miss,
the
requested
address
will
\textit{always}
be
assigned
to
an
empty
cache
entry,
if
one
exists
within
the
dynamic
set
of
the
address.
Only
if
the
dynamic
set
is
completely
filled,
\secache{}
uses
a
random
replacement
policy
to
make
room
for
the
requested
address.
Hence,
contrary
to
other
randomized
cache
architectures,
the
catching
probability
will
be
zero
if
$k'$
does
not
occupy
the
dynamic
set
of
$x$,
which
in
this
case
contains
an
empty
entry.
Since
we
assume
that
the
attack
takes
place
in
a
noise
free
scenario,
we
can
assume
that
addresses
cached
prior
to
the
attack
will
quickly
be
eliminated
based
on
timing.
Only
a
few
very
frequently
accessed
addresses
by
other
processes
will
remain
in
the
cache.
If
the
attacker
relies
on
these
addresses
from
other
processes,
these
addresses
can,
by
the
rules
of
transitivity,
be
considered
part
of
$k'$.

The
dynamic
set
of
$x$
is
determined
by
the
addressing
function
and
in
our
attacker
model
is
assumed
to
be
indistinguishable
from
truly
random.
Hence,
the
probability
that
after
filling
the
cache
with
$|k'|$
non-conflicting
entries,
the
dynamic
set
of
$x$
is
completely
filled
follows
the
hypergeometric
distribution
and
calculates
as

\begin{equation}
p_c(k')
=
\frac{\binom{|k'|}{w}}{\binom{N}{w}}.
\end{equation}

Since
the
attack
\textit{catches}
the
conflict
if
and
only
if
the
dynamic
set
of
$x$
is
filled,
the
above
probability
is
also
the
catching
probability
of
the
profiling
step.

Another
way
to
think
about
this
problem
is
that
every
address
of
the
probing
set
$k'$
taints
the
entry
in
which
it
resides.
All
cache
entries
that
are
empty
remain
untainted.
Since
the
addressing
scheme
assigns
random
entries
to
the
addresses
of
$k'$,
we
receive
a
cache
that
is
partitioned
to
$|k'|$
tainted
entries
at
random
positions
and
the
remaining
$N-|k'|$
untainted
entries.
Importantly,
due
to
the
random
placement,
these
entries
are
equally
distributed
across
the
cache
ways.
The
access
to
address
$x$
selects
a
random
entry
from
each
way,
which
-
due
to
the
equal
distribution
of
tainted
and
untainted
addresses
-
can
be
considered
as
choosing
$w$
random
entries
from
the
cache.
Only
if
all
$w$
chosen
entries
are
tainted,
is
the
profiling
successful.

\begin{table}
\centering
\caption{\label{tab:attack_perf}Size of profiling set $k'$ required to observe an eviction from accessing $x$ with probability $p_c$ in a 16-way cache with 8\;MiB  (131,072 entries). Both schemes use a random replacement policy. Resulting cache utilization by $k'$ in parentheses.}
\footnotesize
\begin{tabular}{c|c|c}
$p_c$	&
\secache{}
	&
\textsc{ScatterCache}~\cite{werner2019scattercache}
\\\hline
1\%
&
98,312
(75.0\%)
&
1,311
(1.0\%)\\
50\%
&
125,520
(95.8\%)
&
65,536
(50.0\%)
\\
90\%
&
130,216
(99.3\%)
&
117,965
(90.0\%)
\\
95\%
&
130,656
(99.7\%)
&
124,519
(95.0\%)
\\
\end{tabular}
\end{table}

Tab.~\ref{tab:attack_perf}
shows
the
size
of
$k'$
required
to
observe
an
eviction
by
the
access
to
the
victim
address
$x$
with
probability
$p_c$
in
a
16-way
cache
for
solely
randomization-based
caches
like
\textsc{ScatterCache}~\cite{werner2019scattercache}
compared
to
\secache{}.
The
results
for
\textsc{ScatterCache}
are
calculated
using
the
findings
by
Purnal
\textit{et
al.}~\cite{purnal2021systematic}.
The
results
show
that
the
attacker
requires
a
much
larger
profiling
set
to
achieve
a
decent
catching
probability
with
\secache
.

\noindent\textbf{Time-Based Evictions (\ding{183})}
During
the
profiling,
addresses
in
$k'$
are
selectively
narrowed
down
from
the
initial
set
of
random
addresses
$k$.
As
discussed
in~\ding{182},
achieving
a
reasonable
catching
probability
requires
filling
well
over
half
of
the
cache
storage.
The
attacker
must
aim
to
minimize
the
time-based
evictions
and
therefore
avoid
conflicts
unrelated
to
the
target
address.
This
keeps
$R_{TTL}$
low
and
thus
enables
the
attacker
to
make
useful
observations
during
the
probe
phase.
For
an
initial
priming
set
$k$,
the
expected
number
of
conflicts
during
the
prime
and
prune
can
be
computed
using
Eq.
\ref{eq:k}.

\begin{equation}
\label{eq:k}
\mathbb{E}[\#c] \geq \sum_{i=1}^{|k|} \sum_{j=1}^{\infty} \left( \frac{\binom{i}{w}}{\binom{N}{w}} \right)^j
\end{equation}

The
amount
of
conflicts
increases
exponentially
with
the
size
of
$k$.
Using
the
above
example
cache
parameters,
priming
the
cache
to
a
50\%
catching
probability
of
the
target
access
would
require
more
than
125,000
accesses
and
result
in
approximately
4,682
conflicts
during
the
priming.
In
the
original
\textsc{Prime+Prune+Probe}
attack,
the
attacker
starts
by
accessing
the
entire
set
of
addresses
during
the
prime
step.
Subsequently,
the
set
is
pruned
by
re-accessing
it
until
no
more
cache
misses
occur.
However,
the
vast
number
of
conflicts
produced
by
priming
the
cache
would
increase
$R_{TTL}$
to
a
level
where
entries
are
evicted
by
timing
rapidly,
undoing
any
progress
made.
Hence,
a
better
approach
is
to
do
the
prime
and
prune
step
incrementally.
Therefore,
the
attacker
adds
addresses
to
$k$
until
a
conflict
occurs.
Then,
they
re-access
all
current
addresses
from
$k$
which
reduces
the
probability
of
time-based
evictions
in
$k$
and
gives
time
for
$R_{TTL}$
to
recover.
In
the
following
we
assume
that
the
$R_{TTL}$
increase
caused
by
one
cache
conflict
can
be
compensated
by
re-accessing
the
primed
addresses
once.
In
reality,
further
accesses
may
be
necessary.

If
we
assume
that
the
attacker
managed
to
prime
the
cache
to
a
sufficient
degree
and
the
access
to
$x$
results
in
a
conflict
(increasing
$R_{TTL}$),
the
attacker
must
now
probe
all
addresses
from
$k'$
to
find
the
one
that
was
evicted
by
$x$.
As
discussed
above,
even
though
the
attacker
may
know
\textit{that}
a
conflict
occurred
by
monitoring
$R_{TTL}$,
they
do
not
know
which
address
of
$k'$
was
evicted.
When
the
attacker
finds
the
first
cache
miss
in
$k'$,
they
cannot
know
if
that
address
has
been
evicted
due
to
the
conflict
with
$x$,
or
due
to
an
expired
TTL.
The
attacker
has
no
other
option
than
to
add
all
potential
candidates
to
$G$
and
-
if
required
-
filter
false
positives
once
a
sufficiently
large
$G$
is
constructed.

\noindent\textbf{Eviction Probability (\ding{184})}
Being
able
to
carry
out
cache
attacks
requires
the
attacker
to
find
many
entries
that
can
collide
with
$x$
in
the
cache.
We
now
investigate
how
many
such
addresses
are
required
to
detect
the
victim
access
with
probability
$p_e$.

Each
address
of
$G$
will
occupy
a
conflicting
cache
entry
with
probability
$\frac{1}{w}$.
However,
only
if
the
entire
dynamic
set
of
$x$
is
occupied
after
accessing
all
addresses
from
$G$,
the
access
to
$x$
will
result
in
a
conflict
and
therefore
be
observable.
Thus,
the
probability
calculates
as

\begin{equation}
p_e
=
\left(
1
-
\left(
1-\frac{1}{w}
\right)
^{\frac{|G|}{w}}\right)^w.
\end{equation}

\begin{table}
\centering
\caption{\label{tab:ev_size}Size of the generalized eviction set $G$ required to observe an eviction from accessing $x$ with probability $p_e$ in a 16-way cache. Both schemes use a random replacement policy.}
\footnotesize
\begin{tabular}{c|c|c}
$p_e$	&
\secache{}
	&
\textsc{ScatterCache}~\cite{werner2019scattercache}
\\\hline
1\%
&
344
&
3
\\
50\%
&
784
&
172
\\
90\%
&
1,247
&
571
\\
95\%
&
1,425
&
743
\\
\end{tabular}
\end{table}

Tab.
\ref{tab:ev_size}
lists
the
required
size
of
$G$
to
observe
an
eviction
by
$x$
with
probability
$p_e$
for
a
16-way
cache
and
compares
it
to
pure
randomization
designs
on
the
example
of
\textsc{ScatterCache}~\cite{werner2019scattercache}.

\noindent\textbf{Profiling Performance Estimation.}
We
now
conservatively
estimate
the
time
it
would
take
an
attacker
to
construct
a
generalized
eviction
set
for
\secache\ and
compare
it
to
purely
randomized
caches.
Therefore
we
assume
a
cache
hit
latency
of
$t_{hit}=10\;ns$
and
a
miss
latency
of
$t_{miss}=20\;ns$.
We
assume
that
the
attacker
knows
when
a
conflict
occurred
e.g.
by
using
the
statistical
approach
described
in
Section~\ref{sec:statistic_ttl},
and,
therefore
can
use
the
described
methodology
in
\ding{183}.
Each
access
to
a
new
address
results
in
a
cache
miss
and
is
hence
counted
with
20\;ns.
Accessing
the
$i$-th
new
address
during
the
construction
of
$k$
results
in
a
conflict
with
an
already
cached
address
of
$k$
with
probability
$p_c$.
Furthermore,
accessing
the
address
evicted
by
the
conflict
leads
to
a
new
conflict
with
probability
$p_c$
again,
repeating
the
process.
Hence,
the
probability
for
$n$
conflicts
calculates
as
$p_c+p_c^2+...+p_c^n$.
When
a
conflict
occurs
from
accessing
a
new
address,
the
attacker
re-accesses
all
prior
entries
of
$k$
until
all
addresses
from
$k$,
including
the
one
that
has
been
evicted
by
the
conflict,
are
cached.
For
real
implementations,
the
attacker
would
need
to
further
keep
re-accessing
entries
from
$k$
until
$R_{TTL}$
recovered
to
a
low
level.
Hence,
we
assume
that
in
each
iteration
the
attacker
adds
an
address
to
$k$
and
if
this
access
causes
a
conflict,
the
attacker
re-accesses
$k$
with
which
results
in
$|k|_i-1$
cache
hits
and
one
cache
miss.
We
do
not
consider
the
noise
effects
introduced
by
time-based
evictions
or
noise
by
other
processes
which
would
make
the
profiling
more
complicated
in
reality.
We
assume
that
the
attacker
aims
to
build
an
eviction
set
$G$
with
$p_e=50\%$.
Overall,
we
calculate
the
time
for
one
iteration
of
profiling
as
\begin{equation}
t_{pp}
=
\sum_{i=1}^{k}
\left(
t_{miss}
+
\sum_{j=1}^{\infty}
p_c(i)^j
\cdot
((i-1)\cdot
t_{hit}+t_{miss})\right).
\end{equation}

\noindent We estimate the time to construct $G$ as

\begin{equation}
\label{eq:time_profiling}
t_k
=
|G|
\cdot
\frac{1}{p_c(k)}
\cdot
t_{pp}.
\end{equation}

Eq.
\ref{eq:time_profiling}
is
composed
of
the
size
of
$G$
required
for
a
probabilistic
eviction
set
with
eviction
probability
$p_e$,
the
probability
that
a
conflict
is
\textit{caught}
during
profiling
($p_c$),
and
the
time
for
one
round
of
profiling.
Using
this
estimation,
the
minimal
profiling
time
for
a
probabilistic
eviction
set
with
$p_e=50\%$
and
the
8
MB
cache
with
16
ways
for
\secache\ is
4,105\;s
(1.1
hours)
if
$k$
occupies
70\%
of
the
cache.
If
the
attacker
can
only
fill
up
to
50\%
of
the
cache,
the
construction
of
$G$
takes
69,575\;s
(19.3
hours).
If
the
attacker
attempts
to
fill
more
than
70\%
of
the
cache,
the
conflicts
will
lead
to
a
higher
profiling
time.
The
same
estimation
for
\textsc{ScatterCache}
with
random
replacement
policy
yields
a
profiling
time
of
0.45\;s
at
1\%
cache
utilization
by
$k$
(a
small
$k$
reduces
the
impact
of
conflicts
during
pruning).
There
are
no
directly
comparable
numbers
available
for
the
recently
proposed
Mirage~\cite{mirageGuru}
and
careful
analysis
is
required
to
assess
the
security
of
Mirage
against
\textsc{Prime+Prune+Probe}.
However,
Mirage~\cite{mirageGuru}
is
similar
to
\secache\ in
that
the
over
provisioned
tag
store
has
empty
entries
and
thus,
conflicts
are
avoided
most
of
the
time.
Therefore,
the
characteristics
regarding
the
catching
probability
are
similar
in
principle.
Mirage
does
not
have
time-based
evictions
which
allows
for
more
optimal
prime
and
prune
without
re-accessing
entries
from
$k$.
Since
our
method
uses
very
conservative
assumptions,
we
expect
all
designs
to
have
a
much
higher
profiling
time
in
reality.
Our
results
show
that
\secache\ can extend the profiling time by
several
orders
of
magnitude
compared
to
\textsc{ScatterCache}.
Since
\secache\ does
not
introduce
indirection
on
the
cache
lookup
path,
the
lookup
latency
is
only
constrained
by
the
randomization
function
which
is
present
in
all
randomized
caches
including
Mirage
and
\textsc{ScatterCache}.

\section{Implementation}
\label{sec:impl}
\noindent
Our
implementation
of
\secache{}
is
consists
of
two
parts.
First,
we
implement
a
functional
model
of
\secache{}
using
gem5~\cite{DBLP:journals/corr/abs-2007-03152},
which
is
the
most
established
simulator
for
CPU
microarchitectures.
Secondly,
to
understand
the
hardware
cost,
we
take
the
effort
to
simulate
a
proof-of-concept
hardware
design
of
\secache{}
using
65nm
CMOS
technology
in
Cadence
Spectre.

Intuitively,
the
performance
of
\secache{}
heavily
depends
on
time-to-live
range
supported
by
the
implementation.
While
the
TTL
reduction
rate
$\rm
R_{TTL}$
dynamically
adapts
the
time-to-live
to
some
extent,
one
should
analyze
the
average
lifetime
of
cache
entries
in
a
traditional
cache
for
the
given
platform.
\relax
We
analyzed
this
time
using
an
unmodified
gem5
and
different
workloads.
It
shows
that
in
regions
of
heavy
cache
usage,
each
entry
hardly
lives
longer
than
50\,ms
without
being
accessed.
Though
outliers
exist,
we
expect
them
to
hardly
affect
the
performance
since
they
are
rarely
accessed
anyway.
We
provide
more
details
of
the
analysis
in
Appendix~\ref{app:classic-cache-analysis}.
\relax
Following
the
setup
chosen
in
related
work~\cite{werner2019scattercache},
our
simulation
features
a
two
level
cache
hierarchy
where
the
1MB
L2
cache
is
a
\secache{}.
Further
information
on
the
gem5
configuration
is
given
in
Section~\ref{sec:eval}.

\subsection{\emph{gem5} Implementation}
\noindent
For
the
randomization
of
the
addressing
scheme,
we
use
a
round
reduced
version
of
the
lightweight
block
cipher
PRINCE~\cite{DBLP:conf/asiacrypt/BorghoffCGKKKLNPRRTY12}
with
three
rounds
and
randomize
the
input
in
each
cache
way
by
xor-ing
a
way-specific
64-bit
secret
to
the
input
of
the
cipher.
We
further
sample
the
bits
used
for
the
cache
index
from
the
entire
output
of
the
randomization
function
instead
of
using,
e.g.,
the
least
significant
bits
as
motivated
in
the
security
analysis
of
the
index-randomization
function
in
Appendix~\ref{sec:sec_addr}.
Since
PRINCE
operates
on
64-bit
blocks,
we
use
the
entire
memory
address
with
zeroed
offset
bits
as
input,
and
therefore
get
a
64-bit
output.
The
bits
used
for
indexing
do
not
need
to
be
stored
as
they
are
implicitly
indicated
by
the
location.
However,
the
tag
size
in
this
scenario
increases
by
6
bits
due
to
the
obfuscated
offset
bits.
This
overhead
can
be
avoided
by
using
a
format
preserving
encryption
scheme
instead
of
PRINCE
as
discussed
in
Section~\ref{subsec:addressing}.

For
the
TTL,
we
modified
the
base
cache
class
of
gem5
and
implemented
a
counter
for
each
cache
entry.
The
TTL
of
each
entry
is
reduced
based
on
a
periodic
event.
Whenever
a
conflict
occurs
--
i.e.,
a
new
address
cannot
be
cached
without
evicting
existing
data
--
the
event
reducing
the
TTL
for
each
entry
is
immediately
triggered.
The
time
until
the
next
event
is
quartered.
Every
time
the
event
executes
without
a
conflict,
the
time
until
the
next
event
scheduling
is
slightly
increased
by
adding
a
constant
value.
This
results
in
a
behavior
as
described
for
$R_{TTL}$
(c.f.
Fig.
\ref{tikz:frequency}).

\subsection{Efficient Hardware Implementation}
\label{subsec:hw_efficient}

\begin{figure}[ht!]
\centering
\includegraphics[width=0.9\linewidth]{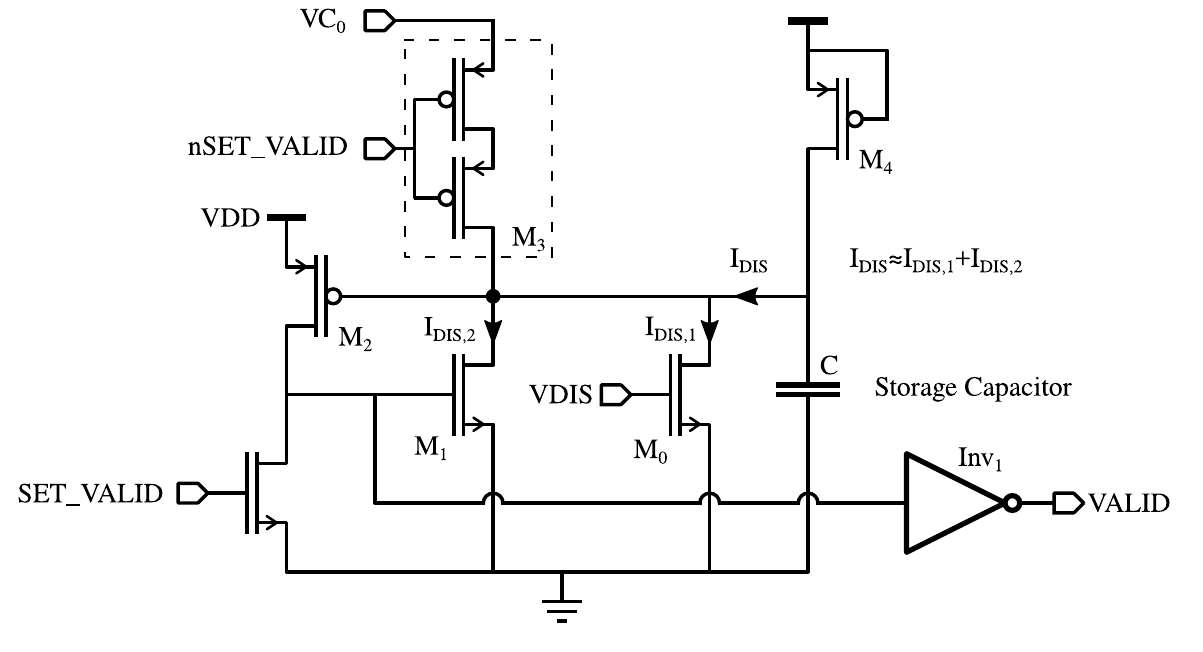}
\caption{Schematic of the proposed delay element.}
\label{fig:Delay_Schematic}
\end{figure}

\begin{figure}[ht!]
\centering
\includegraphics[width=.8\linewidth]{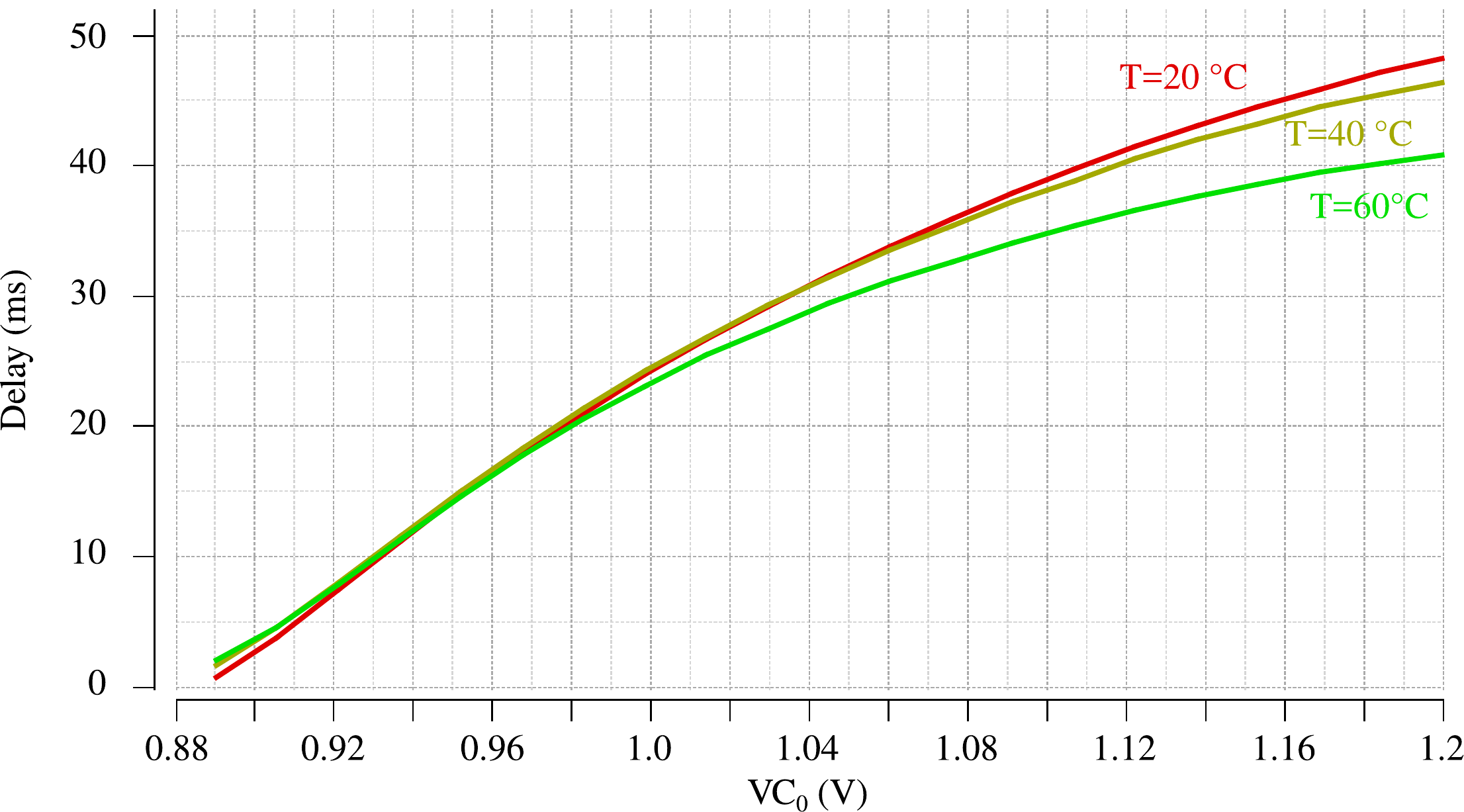}
\caption{Simulated delay of the proposed delay element as a function of the initial capacitor voltage $\rm VC_0$. VDIS was set to 0\,V in this simulation.}
\label{fig:VC0_vs_Delay}
\end{figure}

\begin{figure}[ht!]
\centering
\includegraphics[scale=.9]{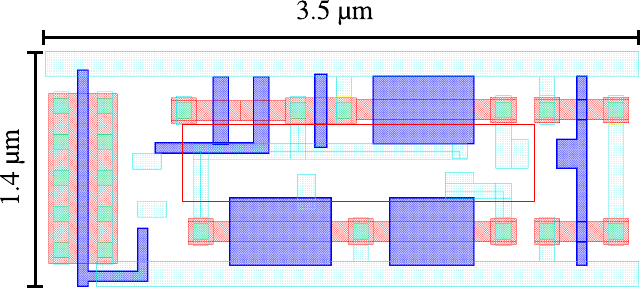}
\caption{Layout of the proposed delay cell (excluding the MIM capacitor).}
\label{fig:Layout}
\end{figure}

\noindent
In
this
section,
we
introduce
a
proof-of-concept
analog
delay
circuit
that
demonstrates
the
feasibility
of
highly
efficient
implementation
of
the
TTL
mechanism.
In
general,
the
placement
of
analog
elements
on
digital
circuitry
requires
careful
decoupling,
for
example
through
shielding
by
ground
planes
and
separating
the
supply
voltage.
Since
these
challenges
are
highly
dependent
on
the
technology
and
layout,
they
remain
for
the
CPU
developer.
It
is
also
possible
to
implement
the
TTL
mechanism
using
digital
counters
for
each
entry.
However,
the
analog
solution
is
more
area
preserving
as
it
is
much
simpler.
We
have
designed
and
simulated
the
analog
delay
element
in
a
65-nm
CMOS
technology.
The
delay
element
is
able
to
achieve
TTLs
of
up
to
50\,ms
(in
terms
of
transistor
switching
speed).
The
basic
idea
behind
the
proposed
delay
element
is
to
charge
a
capacitor
to
an
initial
value
$\rm
VC_0$
that
determines
the
TTL
and
to
discharge
it
by
a
defined
current
$\rm
I_{DIS}$
until
a
threshold
voltage
is
reached.
We
expect,
that
a
metal-insulator-metal
(MIM)
capacitor
with
a
capacitance
of
up
to
100\,fF
can
be
integrated
within
the
metal
stack
over
each
cache
entry
without
competing
for
silicon
area.
The
required
maximum
delay
time
of
50\,ms
consequently
demands
a
very
low
current
of
$\rm
I_{DIS}<1\,pA$.
Such
small
currents
can
be
generated
in
an
area-efficient
manner
by
making
use
of
the
leakage
current
of
transistors,
that
are
either
completely
off
(gate
to
source
voltage
=
0\,V)
or
operate
in
the
deep
sub-threshold
region
with
very
low
gate
to
source
voltages.
The
schematic
of
the
proposed
delay
element
is
shown
in
Fig.~\ref{fig:Delay_Schematic}.
The
signal
SET\_VALID
is
equivalent
to
the
address
line
of
the
cache
entry.
When
the
entry
is
selected,
SET\_VALID
is
high,
while
the
inverted
signal
nSET\_VALID
is
low.
As
a
consequence,
the
storage
capacitor
C
is
charged
to
$\rm
VC_0$
via
transistor
$\rm
M_3$.
The
value
of
$\rm
VC_0$
determines
the
delay
time
of
the
delay
element
and
thus
the
TTL
of
the
cache
entry.
$\rm
M_3$
is
implemented
by
two
series
connected
transistors
by
a
technique
called
stack
forcing
in
order
to
reduce
unwanted
leakage
currents
through
$\rm
M_3$
which
would
otherwise
affect
the
delay
time.
\\After
the
cache
entry
is
deselected
(SET\_VALID
=
0),
C
is
discharged
by
the
leakage
currents
of
$\rm
M_0$
and
$\rm
M_1$.
The
global
low
voltage
signal
VDIS
($\rm
\ll
100\,mV$)
is
used
to
adjust
the
discharge
current
through
$\rm
M_0$.
It
is
used
to
control
the
TTL-reduction
rate
$\rm
R_{TTL}$.
\\
We
found
that
the
addition
of
$\rm
M_4$
improves
the
performance
of
the
delay
element
in
two
ways.
First,
it
increases
the
delay
time
without
increasing
the
area
of
the
capacitor
and,
second,
it
allows
to
compensate
for
the
temperature
variation
of
the
leakage
currents
of
$\rm
M_0$
and
$\rm
M_1$
if
all
contributing
transistors
are
carefully
parameterized.
$\rm
M_1$
and
$\rm
M_2$
form
a
positive
feedback
loop
referred
to
as
a
pseudo-CMOS-thyristor\cite{Kim_ThyristorDelay}.
After
C
is
discharged
below
the
threshold
voltage
of
$\rm
M_2$,
$\rm
M_2$
charges
the
gate
of
$\rm
M_1$
which
then
becomes
conductive
and
discharges
the
remaining
charge
of
C
in
$\rm
\approx
10\,ns$.
The
advantage
of
this
feedback
loop
is
the
drastic
reduction
of
short
circuit
current
in
the
output
inverter
$\rm
Inv_1$,
which
would
be
severe
if
the
complete
discharge
happened
on
a
ms-timescale.
\\
The
simulated
delay
time
of
this
circuit
as
a
function
of
$\rm
VC_0$
is
shown
in
Fig.~\ref{fig:VC0_vs_Delay}
for
three
different
temperatures.
For
$0.89\,V
<
\rm
VC_0
<
1.2\,V$
delay
times
between
1\,ms
and
almost
50\,ms
are
achieved.
It
should
be
noted
that
the
observed
moderate
variation
with
temperature
is
actually
beneficial,
as
it
adds
randomness
to
the
TTL.
\\To
estimate
the
area
required
to
implement
this
circuit,
we
have
designed
the
layout
shown
in
Fig.~\ref{fig:Layout}.
With
an
area
of
$\rm
3.5\,\mu
m
\times
1.4\,\mu
m
=
4.9\,\mu
m^2
$
we
can
estimate
the
area
overhead
of
this
circuit:
Assuming
the
size
of
an
SRAM
cell
in
65-nm
CMOS
technology
to
be
$\rm
0.5\,\mu
m^2$
as
reported
in~\cite{Ohbayashi_SRAM_Cell},
the
area
of
a
128
bit
cache
entry
(64
bit
data,
$\approx$
60
bit
tag
plus
some
flags)
is
$\rm
64\,\mu
m^2$
(without
overhead).
We
therefore
estimate
the
area
overhead
of
the
analog
TTL
implementation
circuit
to
be
$<
8\%$.
In
the
simulation
we
used
a
MIM
capacitor
with
an
area
of
$\rm
30\,\mu
m^2$
which
can
be
conveniently
placed
in
the
metal
stack
above
the
cache
entry.
\\
If
the
charge
time
of
the
capacitor
$\rm
t_c$
is
too
long,
it
may
increase
the
access
time
of
the
cache.
We
have
therefore
simulated
how
fast
the
capacitor
is
charged.
The
result
of
our
simulations
was
that
$\rm
t_c$
depends
on
$\rm
VC_0$.
For
$\rm
VC_0
=
1.2\,V$
the
recharge
time
was
8\,ns
while
for
the
minimal
value
of
$\rm
VC_0
=
0.89\,V$,
$\rm
t_c$
was
41\,ns.
This
corresponds
well
to
the
reported
L3
cache
latency
of
an
i7
CPU\cite{levinthal2009performance}
of
40
to
300
cycles
(13\,ns
to
100\,ns
with
a
clock
frequency
of
3\,GHz).
\\The
generation
of
the
analog
control
signals
$\rm
VC_0$
and
$\rm
VDIS$
is
not
within
the
scope
of
this
work
since
many
suitable
digital-to-analog
converters
(DAC)
have
been
published.
DACs
can
be
realized
with
a
small
area
and
power
consumption.
E.g.
in~\cite{Kwon_DAC}
a
current
steering
DAC
with
an
area
consumption
of
$\rm
0.1\,mm^2$
in
65-nm
CMOS
technology
and
a
power
consumption
of
12\,mW
is
reported.

\section{Evaluation}
\label{sec:eval}
\noindent
We
evaluate
\secache{}
using
the
gem5
CPU
simulator~\cite{DBLP:journals/corr/abs-2007-03152}.
In
order
to
compare
the
performance
of
our
new
cache
concept
we
execute
popular
benchmarks
including
MiBench~\cite{990739},
Parsec~\cite{bienia2008parsec}
and
SPEC
CPU
2017~\cite{BibEntry2021Mar}.
We
provide
a
detailed
performance
evaluation
of
\secache{}
and
compare
it
to
traditional
caches.
We
execute
the
MiBench
and
Parsec
benchmark
using
gem5's
full
system
mode
to
get
the
most
precise
performance
result.
We
simulate
Ubuntu
18.04
LTS
with
kernel
4.19.83.
For
the
SPEC
CPU
2017
Benchmark
we
use
gem5's
Syscall
Emulation
(SE)
mode.
Due
to
the
vast
resource
usage
and
high
simulation
time
it
is
not
feasible
to
use
full
system
mode.
The
SE
mode
also
provides
high
accuracy
in
performance
measurements,
but
omits
the
simulation
of
the
entire
operating
system.
Other
related
work
often
only
chooses
representative
code
slices
of
the
benchmark~\cite{werner2019scattercache,Tan2020}.

For
the
purpose
of
the
evaluation
we
use
gem5's
default
cache
implementation
as
reference,
hereinafter
referred
to
as
\textit{classic}.
We
focus
on
the
comparison
of
\secache{} and classic caches since the most related concepts, e.g., \textsc{ScatterCache}~\cite{werner2019scattercache} are close to the classic cache performance. As discussed earlier in this paper, Mirage~\cite{mirageGuru} changes the critical path of the cache
access
and
may
therefore
induce
additional
overhead.
In
\secache,
the
access
latency
is
only
affected
by
the
randomization
function
which
must
be
low-latency
in
order
to
avoid
delays.
Following
related
work~\cite{werner2019scattercache},
both
the
classic
cache
and
\secache{}
have
a
two
level
cache
hierarchy
with
a
L1
and
and
a
mostly
inclusive
L2
cache.
Based
on
our
observations
regarding
the
average
lifetime
of
cache
entries
(Section~\ref{sec:impl}),
we
have
chosen
a
maximum
TTL
of
50ms
for
the
evaluation.
We
describe
the
simulation
details
including
cache
associativity
and
choosen
benchmark
workloads
in
Appendix~\ref{app:configuration}.

\relax

\begin{figure}[ht!]
	\centering
	\includegraphics[width=.8\linewidth]{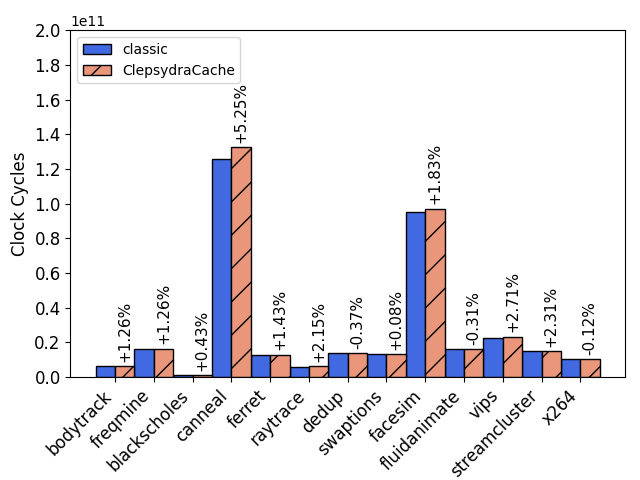}
		\caption{\label{fig:parsec:totalclocks}
Comparison
of
clock
cycles
for
programs
from
the
Parsec
benchmark
suite.}
\end{figure}

\noindent\textbf{Performance.}
We
use
clock
cycles
as
a
metric
to
measure
the
absolute
performance
of
the
benchmarks.
For
the
Parsec
benchmark
suite,
the
overall
performance
results
are
shown
in
Fig.~\ref{fig:parsec:totalclocks}.
Overall,
the
performance
of
\secache{}
matches
the
performance
of
traditional
caches
with
a
penalty
between
-0.37\%
and
+5.25\%.
On
average
the
performance
penalty
is
at
1.38\%.

\begin{figure}[ht!]
	\centering
	\includegraphics[width=.8\linewidth]{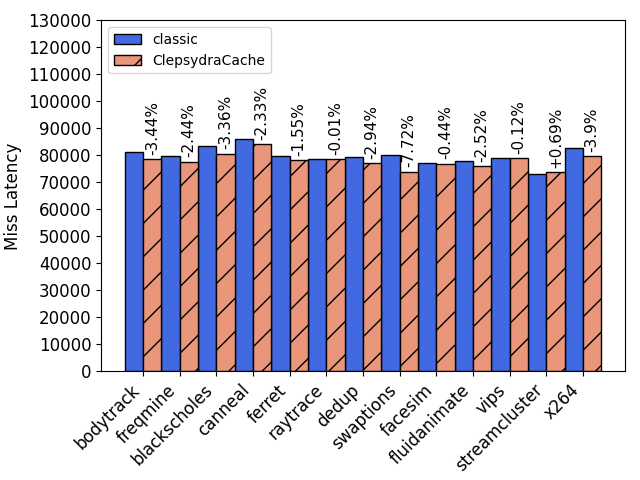}
		\caption{\label{fig:parsec:avgmisslat}
Comparison
of
the
average
miss
latency
for
the
programs
from
the
Parsec
benchmark
suite.}
\end{figure}

While
the
fact
that
\secache{}
improves
the
performance
for
some
benchmarks
appears
counterintuitive
at
first
glance,
the
reason
for
this
can
be
found
by
looking
at
the
average
miss
latency
depicted
in
Fig.~\ref{fig:parsec:avgmisslat}.
It
shows
that
with
the
exception
of
the
\textit{streamcluster}
benchmark,
\secache{}
consistently
lowers
the
average
miss
latency
and
thus,
compensates
for
the
slightly
increased
miss
rate.
The
reason
for
that
can
be
found
in
the
writeback
of
dirty
cache
entries.
While
a
traditional
cache
performs
writebacks
on
modified
data
when
a
conflict
occurs,
\secache{}
performs
writebacks
based
on
timing
and
thus
is
independent
of
any
cache
misses.
Hence,
the
miss
latency
on
traditional
caches
includes
the
time
needed
for
a
potential
writeback
whereas
\secache{} usually invalidates and writes back the entries
before
a
conflict
occurs.
Therefore,
nearly
all
benchmarks
have
a
slightly
lower
average
miss
latency.
The
\textit{streamcluster}
benchmark
is
the
one
outlier
of
this
observation.
The
workload
characteristic
given
in~\cite{bienia2008parsec}
reveals
that
this
benchmark
performs
exceptionally
few
writes
and
hence,
hardly
any
cache
entries
need
to
be
written
back.
Thus,
the
influence
of
this
effect
is
minimized
and
the
occurrence
of
\textit{some} conflict-based evictions leads to a slightly higher average
miss
latency
for
\secache{}.
The
low
overall
miss
latency
of
\textit{streamcluster} supports this finding.

While
the
performance
increase
of
\secache{}
in
gem5
appears
like
a
genuine
improvement
over
traditional
caches,
from
a
security
point
of
view
a
constant
time
cache
would
be
much
preferable.
Though
in
our
simulation
the
timing
difference
between
a
cache
miss
with
and
without
a
writeback
was
sufficiently
small
to
not
be
reliably
measurable,
this
may
not
be
the
case
with
all
caches.
Equally,
it
may
also
be
that
some
cache
implementations
do
not
have
this
timing
difference.
Even
if
we
assume
that
the
miss
latency
is
constant
between
the
classic
cache
and
\secache{}
in
our
evaluation,
the
overall
average
miss
latency
only
increases
by
about
2.3\%
on
average.
This
would
only
marginally
affect
the
performance
data
shown
in
Fig.~\ref{fig:parsec:totalclocks}.

\begin{figure}[t]
	\centering
	\includegraphics[width=.8\linewidth]{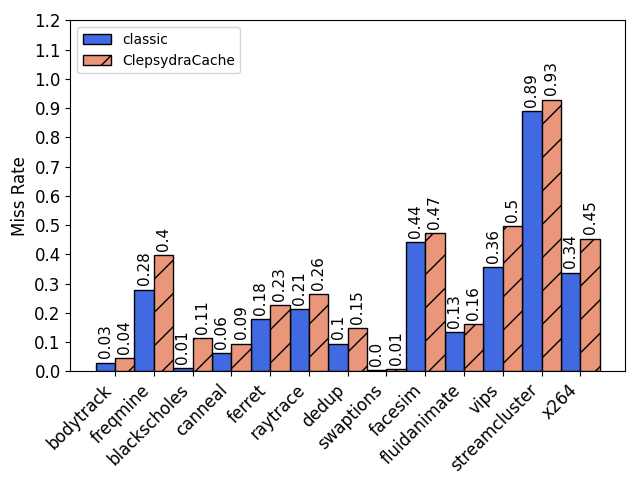}
		\caption{\label{fig:parsec:missrate}
Comparison
of
the
miss
rate
for
the
programs
from
the
Parsec
benchmark
suite.}
\end{figure}

As
one
would
expect,
\secache{}
experiences
a
slightly
increased
miss
rate
compared
to
classic
caches.
That
is,
since
sometimes
entries
are
evicted
based
on
expired
TTLs,
but
are
later
referenced
and
need
to
be
re-loaded.
Fig.~\ref{fig:parsec:missrate}
shows
the
miss
rate
for
Parsec,
i.e.
the
ratio
of
cache
accesses
that
resulted
in
a
cache
miss.
The
hit
rate
is
the
inverted
value
and
calculates
as
$1-
\text{Miss
Rate}$.
The
miss
rate
is
generally
close
to
the
miss
rate
of
traditional
caches.
Benchmarks
with
a
low
miss
rate
for
classic
caches
experience
a
low
miss
rate
for
\secache{}.
Table~\ref{tab:spec_performance}
reports
the
results
for
SPEC
CPU
2017.
The
figures
for
SPEC
are
also
available
in
Appendix~\ref{app:spec}
and
for
MiBench
in
Appendix~\ref{app:mibench}.
Generally,
the
results
using
SPEC
CPU
2017
and
MiBench
are
very
similar
to
the
ones
reported
for
Parsec.

\begin{table}
\centering
\footnotesize
\caption{\label{tab:spec_performance}Benchmark results of the SPEC CPU 2017 benchmark suite using \secache{} in comparison to a traditional cache.}
\begin{tabular}{c | c | c | c | c}
\textbf{Benchmark}& \textbf{Clock} & \textbf{\shortstack[c]{Miss Rate \\(difference)}} & \textbf{\shortstack[c]{Avg. \\Miss Latency}} & \textbf{\shortstack[c]{Conflict\\Evictions}} \\ \hline\hline
Deepsjeng
&-1.72\%&$\pm$0\%&-4.38\%&
-94.13\%\\\hline
Exchange2
&+0.04\%&+8\%&-0.78\%&-100\%\\\hline
Gcc
&+0.39\%&+5\%&-2.87\%&
-95.71\%\\\hline
Leela
&+0.11\%&+2\%&-2.34\%&-95.55\%\\\hline
Perlbench
&$\pm0\%$
&+3\%
&-0.2\%
&
-100\%
\\\hline
x264
&
+0.24\%&+2\%&-1.36\%&-98.93\%
\\\hline
Xalancbmk
&+0.3\%
&+3\%
&-2.35\%
&-98.61\%\\\hline
Xz
&-0.11\%
&+1\%
&-2.83\%
&-95.54\%\\
\end{tabular}
\end{table}

\begin{figure}[ht!]
	\centering
	\includegraphics[width=.8\linewidth]{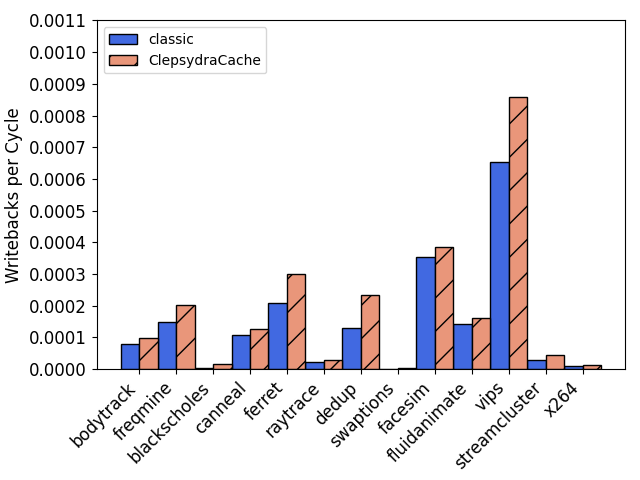}
		\caption{\label{fig:parsec:wb_cycle}
Writebacks
per
cycle
for
\secache{}
using
the
Parsec
benchmark
suite.}
\end{figure}

We
investigate
the
number
of
writebacks
issued
with
and
without
\secache{}
in
Fig.~\ref{fig:parsec:wb_cycle}.
Since
more
entries
are
written
back
using
\secache{},
the
writebacks
per
cycle
are
slightly
increased.
However,
the
main
incentive
for
this
evaluation
was
to
not
overload
the
writeback
buffer.
The
increase
in
writebacks
per
cycle
is
only
minor
and
hence,
hardly
affects
the
size
of
the
hardware
logic
required
for
writebacks
(e.g.
the
writeback
buffer
size).
Finally,
we
compare
the
average
lifetime
of
cache
entries
for
classical
caches
and
\secache{}
in
Appendix~\ref{app:classic-cache-analysis}.
Our
implementation
limits
the
maximum
lifetime
of
an
entry
to
50\,
ms.
This
value
is
chosen
based
on
the
analysis
of
the
average
entry-lifetime
in
a
classic
cache
for
our
setup.
It
shows
that
the
average
lifetime
of
entries
in
\secache{}
is
very
similar
to
the
one
in
classic
caches.

\begin{figure}[ht!]
	\centering
	\includegraphics[width=.8\linewidth]{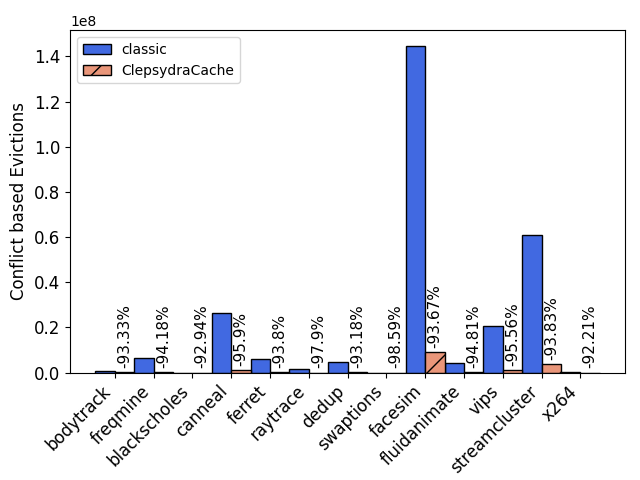}
		\caption{\label{fig:parsec:confevict}
Comparison
of
conflict-based
evictions
for
the
programs
from
the
Parsec
benchmark
suite.}
\end{figure}

\noindent\textbf{Security.}
Fig.~\ref{fig:parsec:confevict}
depicts
the
rate
of
cache
conflicts
in
\secache\ compared
to
traditional
caches
using
the
PARSEC
benchmark
suite.
The
figures
for
the
other
benchmarks
are
shown
in
Appendix
\ref{app:spec}
and
Appendix
\ref{app:mibench}.
Cache
conflicts
are
an
important
occasion
for
an
attacker
to
learn
conflicting
addresses
and
construct
eviction
sets.
With
\secache{},
the
number
of
cache
conflicts
is
reduced
more
than
90\%
for
all
Parsec
benchmarks
(on
average
94.6\%)
which
greatly
reduces
the
chances
of
an
attacker
observing
an
eviction.
As
discussed
in
our
security
analysis,
provoking
a
targeted
cache
conflict
is
very
challenging
with
\secache{}.
Hence,
the
cost
for
a
successful
attack
is
not
proportional
to
the
observed
conflicts.
To
verify
that
our
\secache\ implementation
is
secure
against
\textsc{Prime+Prune+Probe} attacks, we implement the attack
and
execute
it
in
the
gem5
SE
mode.
Using
a
pure
randomized
cache
design
which
mimics
\textsc{ScatterCache}~\cite{werner2019scattercache},
the
construction
of
an
eviction
set
with
$p_e=90\%$
took
2.15
seconds.
Using
\secache\,
we
were
not
able
to
find
conflicting
addresses.
We
aborted
the
attack
without
success
after
77
simulated
seconds
(two
days
of
simulation).
After
that
time,
not
a
single
conflicting
address
has
been
found.
Furthermore,
the
small
timing
difference
between
cache
conflicts
with
and
without
writebacks
was
not
measurable
in
our
experimental
setup.
However,
if
one
were
to
implement
a
side-channel
secure
cache,
such
timing
differences
should
be
avoided.
Note
that
it
is
easily
feasible
to
reduce
the
amount
of
conflict-based
evictions
even
further
by
reducing
the
maximum
time-to-live,
albeit
at
the
cost
of
a
higher
miss
rate
and
therefore,
reduced
performance.

\section{Conclusion}
\label{sec:conclusion}
\noindent
We
presented
\secache{},
a
novel
cache
architecture
that
eliminates
cache
attacks
by
preventing
the
attacker
from
constructing
(generalized)
eviction
sets.
Our
solution
is
purely
architectural
and
therefore
backwards
compatible
to
the
entire
software
stack.
We
are
the
first
to
explore
an
analog
proof-of-concept
hardware
design
in
65nm
CMOS
technology
which
facilitates
an
area
efficient
implementation
of
\secache
.
We
showed
that
\secache{}
has
minimal
performance
overhead
and
in
some
cases
even
outperforms
traditional
cache
architectures
using
representative
workloads.

\section{Acknowledgements}
This
work
is
funded
by
the
Deutsche
Forschungsgemeinschaft
(DFG,
German
Research
Foundation)
under
Germany's
Excellence
Strategy
-
EXC
2092
CASA
-
390781972
and
by
the
DFG
under
the
Priority
Program
SPP
2253
Nano
Security
(Project
RAINCOAT
-
Number:
440059533).
"Any
opinions,
findings,
and
conclusions
or
recommendations
expressed
in
this
material
are
those
of
the
author(s)
and
do
not
necessarily
reflect
the
views
of
the
funding
agencies".
Date
of
this
document:
August
12th,
2022.

\bibliographystyle{plain}
\bibliography{papers.bib}

\appendix

\section{Security: Further Attacks}
\label{app:attacks}

In
this
Appendix,
we
discuss
attack
scenarios
beyond
\textsc{Prime+Prune+Probe}.

\paragraph{Evict+Time Attacks.}
\noindent
In
an
\textsc{Evict+Time}~\cite{Osvik-2006-CacheAttacksandCo}
attack,
the
attacker
measures
the
execution
time
of
a
victim
program
before
evicting
a
specific
cache
set.
The
attacker
then
executes
the
victim
program
again
and
measures
whether
the
timing
differentiates
from
the
first
run.
If
so,
the
attacker
can
conclude
that
the
data
in
the
set
was
accessed
by
the
victim
process
during
execution.
This
attack
does
not
work
with
\secache{}.
Next
to
the
requirement
of
performing
targeted
evictions
of
data
from
a
victim
process,
the
randomization
of
the
TTL
induces
random
deviations
to
the
execution
time
of
the
victim
program.
This
does
not
only
affect
the
targeted
cache
set
but
any
memory
access
performed
during
the
execution
of
the
victim
process.
Hence,
the
attacker
has
no
indication
whether
timing
deviations
in
the
execution
time
of
a
program
are
a
result
of
targeted
evictions
resulting
from
conflicts
or
random
evictions
resulting
from
expired
TTL
values.

\paragraph{Index Randomization.}
\label{sec:sec_addr}
\noindent
The
TTL
feature
in
combination
with
index
randomization
successfully
prevents
a
useful
observation
of
a
cache
conflict.
However,
the
design
of
the
index
randomization
function
described
in
Section~\ref{subsec:addressing}
requires
that
an
attacker
is
not
able
to
construct
additional
conflicting
addresses
from
already
observed
ones.
In
particular,
it
implicitly
requires
that
hardness
of
the
problem
deducing
information
about
the
key
from
the
observation
that
two
addresses
are
conflicting.
As
conflicting
addresses
correspond
to
certain
bits
being
identical
in
the
output
of
the
index
randomization
function,
this
situation
is
similar
to
a
truncated
differential
attack
on
block
ciphers
as
originally
introduced
by
Knudsen~\cite{DBLP:conf/fse/Knudsen94}.
Here
the
attacker
could
in
principle
try
to
guess
part
of
the
first
round
key
and,
based
on
the
guess,
choose
specific
input
differences
that
increase
the
probability
of
the
truncated
differential
and
thus
the
probability
of
creating
conflicting
addresses.
This
implies
that
the
probability
of
the
truncated
differential
should
not
be
too
large.
Note
that
the
attacker
cannot
choose
the
truncated
output
difference
to
consider,
but
this
is
implicitly
defined
by
the
design.
In
particular
one
possibility
to
ensure
small
probability
is
to
spread
the
output
bits
used
for
defining
the
index
across
the
entire
output
of
the
cipher
and
in
particular
across
as
many
S-boxes
as
possible.

The
rather
limited
data
complexity
an
attacker
can
use
in
combination
with
the
high
level
of
noise
created
by
the
TTL
feature
suggest
that
a
reduced
round
version
of
secure
block
cipher
suffices
to
ensure
security.
However,
analyzing
the
details
of
such
an
approach
is
beyond
the
scope
of
this
work.

\paragraph{Denial of Service}
\noindent
Since
the
reduction
rate
$\rm
R_{TTL}$
of
the
TTL
reduction
depends
on
the
number
of
conflicts
that
occurred,
an
attacker
could
effectively
clear
all
cache
entries
by
making
a
huge
amount
of
memory
accesses
to
uncached
addresses.
This
behavior
would
cause
frequent
conflicts
and
hence
increase
the
reduction
rate,
evicting
all
entries
in
a
short
time
period.
The
attacker
could
further
keep
accessing
new
memory
addresses
and
hence,
effectively
disable
the
cache
as
new
entries
are
deleted
relatively
fast
due
to
the
high
TTL
reduction
rate.
In
a
traditional
cache,
the
attacker
could
achieve
a
similar
behavior
by
accessing
many
uncached
memory
addresses
and
therefore
evicting
existing
entries
from
other
processes.
Importantly,
no
information
can
be
leaked
during
a
DoS
attack.

\section{Configuration Details}
\label{app:configuration}

The
MiBench
and
Parsec
benchmark
are
executed
both
in
gem5's
full
system
mode
which
simulates
all
of
the
hardware
from
the
CPU
to
the
I/O
devices
and
gives
a
precise
performance
result.
We
simulate
Ubuntu
18.04
LTS
with
kernel
version
4.19.83.
However,
due
to
the
vast
resource
usage
and
the
corresponding
simulation
time,
it
is
not
feasible
to
run
a
full
system
simulation
of
SPEC
CPU
2017.
While
related
work
mostly
reduced
the
workload
of
SPEC
by
choosing
representative
code
slices
from
the
benchmark~\cite{werner2019scattercache,Tan2020},
we
simulate
the
SPEC
Integer
Speed
suite
based
on
the
command
line
interface
described
in~\cite{Limaye2018}
using
gem5's
Syscall
Emulation
(SE)
mode
which
provides
accurate
performance
simulation
of
the
benchmark
but
omits
simulating
the
operating
system.
To
further
reduce
the
simulation
time,
we
use
the
\textit{test}
workloads
for
SPEC
which
in
many
cases
feature
smaller
input
sizes
but
generally
perform
the
same
tasks
as
the
\textit{reference}
workloads.
For
the
Parsec
benchmarks
we
use
the
input
size
\textit{simmedium}
and
for
MiBench
the
input
size
\textit{large}.

For
the
simulated
system
we
have
chosen
a
setup
with
\SI{4}{\gibi\byte}
of
RAM
and
a
clock
speed
of
\SI{2}{\GHz}.
The
L1
data
and
the
instruction
caches
in
both
scenarios,
i.e.
the
classic
gem5
reference
and
\secache{},
are
modeled
as
4-way
set
associative
classic
cache
with
a
size
of
\SI{64}{\kibi\byte}.
The
L2
cache,
which
is
in
our
case
also
the
LLC,
has
a
size
of
\SI{1}{\mebi\byte}
and
is
8-way
set
associative
in
both
cases.
We
apply
the
new
\secache{}
architecture
to
the
LLC,
i.e.
L2,
and
use
a
Random-Replacement-Policy
(RRP).
For
the
classic
caches
we
use
LRU~\cite{DBLP:conf/ACMse/Al-ZoubiMM04}
replacement
policy.
We
provide
an
overview
of
the
cache
configuration
of
our
setup
in
Table~\ref{tab:sumCacheSettings}.

\begin{table}[ht!]
\centering
\footnotesize
\caption{\label{tab:sumCacheSettings} Summary of cache settings in gem5's simulation environment.}
\begin{tabular}{l|cc}
Settings
&
Classic
&
\secache{}
\\
\hline
L1
Size
&
64kB
&
64kB
\\
L1
Architecture
&
classic
&
classic
\\
L1
Replacement
Policy
&
LRU
&
LRU
\\
L2
Size
&
1MB
&
1MB
\\
L2
Architecture
&
classic
&
\secache{}
\\
L2
Replacement
Policy
&
LRU
&
Random
Replacement
\end{tabular}
\end{table}

\section{Analysis of traditional caches}
\label{app:classic-cache-analysis}

\begin{figure}[ht!]
	\centering
	\includegraphics[width=.9\linewidth]{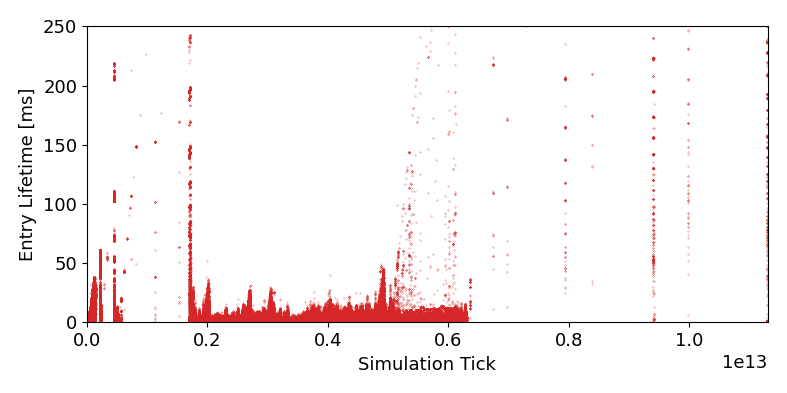}
		\caption{\label{fig:l2lifetime}
Lifetime
of
L2
cache
entries
during
system
boot,
Parsec's
\textit{freqmine}
benchmark
and
shutdown
using
traditional
caches.
Every
red
dot
indicates
the
lifetime
of
an
entry
that
has
been
evicted.
Lifetime
normalized
by
(hit)
accesses.}
\end{figure}

\begin{figure}[ht!]
	\centering
	\includegraphics[width=.9\linewidth]{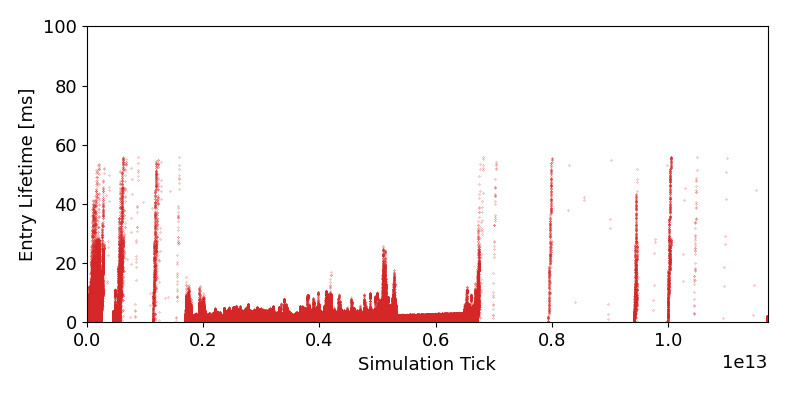}
		\caption{\label{fig:ttllifetime}
Lifetime
of
L2
\secache{}
entries
during
system
boot,
Parsec's
\textit{freqmine}
benchmark
and
shutdown.
Every
red
dot
indicates
the
lifetime
of
an
entry
that
has
been
evicted.
Lifetime
normalized
by
(hit)
accesses.}
\end{figure}

In
order
to
choose
an
upper
limit
for
our
TTL
on
our
reference
setup
(see
Section~\ref{sec:impl}),
we
have
evaluated
the
average
lifetime
of
an
entry
in
the
cache
during
boot
and
normal
operation.
An
example
of
a
boot,
followed
by
a
benchmark
is
given
by
figure~\ref{fig:l2lifetime}
and
figure~\ref{fig:ttllifetime}.

As
seen
in
figure~\ref{fig:l2lifetime}
most
cache
entries
have
a
lifetime
shorter
than
50ms
in
our
classic
cache
setup.
We
have
similar
results
for
other
workloads.
Therefore,
we
have
chosen
50ms
as
the
upper
TTL
limit
for
our
implementation
and
evaluation
as
seen
in
figure~\ref{fig:l2lifetime}.
The
choice
of
the
upper
TTL
bound
will
affect
performance
and
needs
to
be
re-evaluated
for
other
processor
and
cache
setups.

\section{\secache\ and Real-Time-Systems}
\label{app:real-time}

Systems
can
have
different
requirements
towards
expected
or
mandatory
response
times.
In
case
a
system
requires
strict
deadlines
towards
response
times,
it
is
called
a
real-time
system.
There
are
different
types
of
real-time
systems:
\textit{hard,
soft
and
firm
real-time
systems}.
Those
types
can
be
differentiated
by
their
strictness
towards
deadlines.
A
hard
real-time
system
must
meet
its
deadlines
at
all
times.
Otherwise,
this
can
lead
to
severe
consequences.
A
good
example
for
hard
real-time
systems
are
control-units
for
airbags.
In
contrast,
a
firm
real-time
system
can
accept
infrequent
deadline
misses.
Voice
transmission
is
such
a
system.
A
few
missed
bits
of
sound
will
not
degrade
the
entire
system,
but
frequent
misses
will.
A
real-time
system
which
is
neither
hard
nor
firm,
is
called
soft.
In
a
soft
real-time
system
the
value
of
the
information
decreases
over
time.
A
good
example
is
an
air
conditioning
system
with
regular
temperature
measurements.
The
air
conditioner
operates
on
the
most
frequent
measurements.
Frequent
misses
will
be
tolerated
as
long
as
there
are
a
few
timely
measurements
available~\cite{Shin94realtimecomputing}.
For
a
hard
real-time
system,
it
is
very
important
to
be
able
to
predict
a
worst-case
time
to
complete
its
deadlines.

Thus,
for
the
compatibility
of~\secache\ with
real-time
systems,
the
worst-case
delay
of~\secache\ needs
to
be
considered.
The
worst-case
delay
is
known
and
represents
the
case
with
no
cached
entries.
We
assume
for
the
worst-case
delay,
that
all
cached
entries
are
evicted
due
to
an
expired
TTL.
Thus,
like
for
all
other
caches
the
(worst-case)
influence
of~\secache\ on
the
execution
time
is
precisely
known.
In
general
there
is
no
conceptual
difference
between
a
classic
cache
and~\secache\ as
in
both
scenarios
the
worst-case
scenario
i.e.
no
cached
entries
needs
to
be
considered
by
the
real-time
system.
The
choice
of
the
cache
influences
the
time
required
for
the
\textit{CPU}
to
perform
\textit{load}
and
\textit{store}
instructions.
For
example,
we
consider
a
hard
real-time
system
with
strict
response
times
$t_{response}$.
The
system
needs
to
have
enough
computing
power
$p$
to
complete
the
task
$tsk$
before
$t_{response}$
(deadline).
A
designer
of
a
hard
real-time
system
always
chooses
enough
power
$p$
to
complete
$tsk$
before
$t_{response}$
under
worst-case
circumstances.
As
a
consequence,
there
is
no
difference
for
the
choice
of
a
cache
on
a
hard
real-time
system.

In
case
of
a
firm
or
soft
real-time
system
an
average
response
time
$t_{avg}$
for
a
deadline
$d$
can
be
considered.
Since
a
few
missed
deadlines
are
allowed
in
firm
or
soft
real-time
systems,
the
system
could
be
designed
to
operate
on
average
execution
times.
The
choice
of
the
cache
will
now
influence
the
(average)
execution
of
\textit{load}
or
\textit{store}
instructions
on
the
\textit{CPU}.
However,
different
caches
will
lead
to
different
average
execution
times
for
different
use-cases.
Thus,
the
design
of
such
a
real-time
system,
needs
to
determine
the
average
execution
times
for
the
used
cache
$c$
with
workload
$w$.
Therefore,
the
only
difference
between
a
classic
cache
and~\secache\ is
the
average
execution
time
under
a
specific
workload
$w$.
\section{SPEC Benchmark Results}
\label{app:spec}
This
Appendix
shows
the
graphs
for
the
SPEC
CPU
2017
benchmarks.
The
benchmarks
\textit{omnetpp}
and
\textit{mcf}
did
not
finish
simulation
in
a
reasonable
time
frame
and
are
hence
left
out.

\begin{figure}[H]
	\centering
	\includegraphics[width=\linewidth]{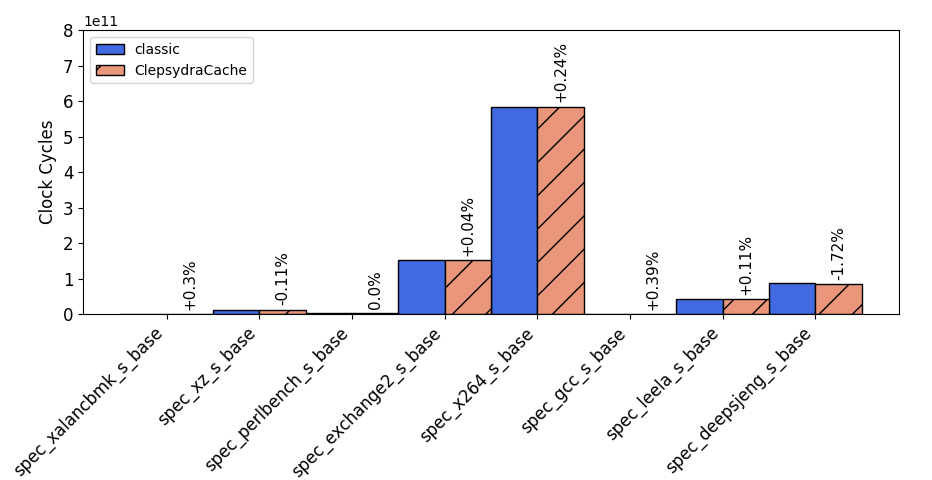}
		\caption{\label{fig:spec:clock}
Comparison
of
runtime
for
the
programs
from
the
SPEC
CPU
2017
benchmark
suite.}
\end{figure}

Fig.
\ref{fig:spec:clock}
shows
the
elapsed
clock
cycles
for
the
SPEC
CPU
2017
benchmark
suite
with
traditional
caches
vs.
\secache{}.
The
overhead
falls
in
the
range
between
-1.72\%
and
+0.39\%
with
an
average
of
-0.07\%.
Thus,
\secache{}
slightly
outperforms
traditional
caches
using
the
SPEC
CPU
2017
benchmarks.
As
with
the
Parsec
benchmarks
evaluated
in
the
main
paper,
this
speedup
can
be
traced
to
a
reduced
average
miss
latency
due
to
less
writebacks
of
dirty
entries
on
access.

\begin{figure}[H]
	\centering
	\includegraphics[width=\linewidth]{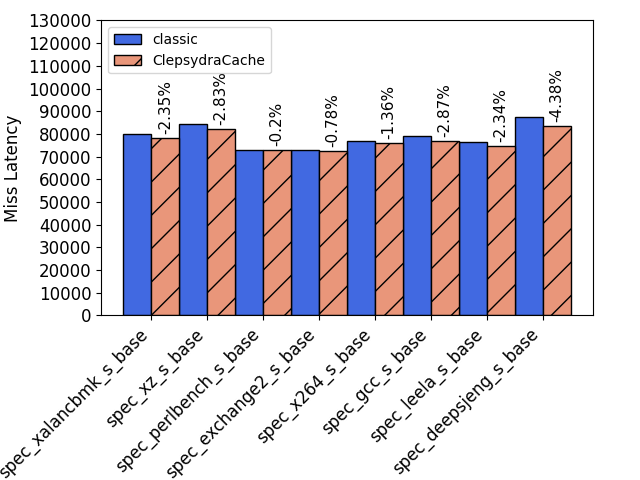}
		\caption{\label{fig:spec:misslat}
Comparison
of
the
average
miss
latency
for
the
programs
from
the
SPEC
CPU
2017
benchmark
suite.}
\end{figure}

Fig.
\ref{fig:spec:misslat}
shows
the
difference
in
average
miss
latency
for
the
SPEC
CPU
2017
benchmark
suite
using
traditional
caches
vs.
\secache{}.
The
average
miss
latency
reduces
by
between
0.2\%
and
4.38\%
for
the
SPEC
benchmarks.

\begin{figure}[H]
	\centering
	\includegraphics[width=\linewidth]{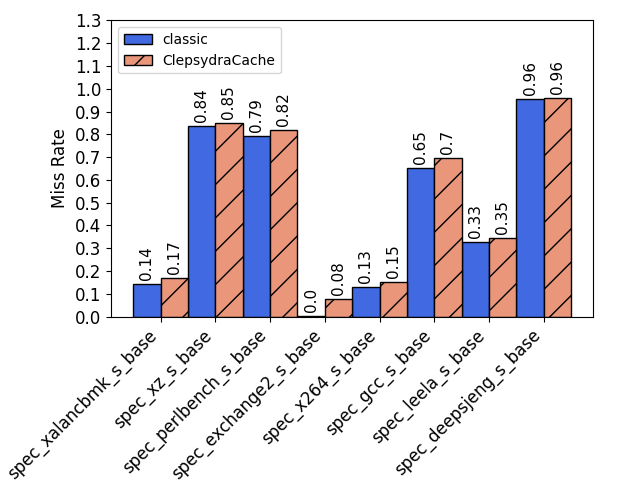}
		\caption{\label{fig:spec:missrate}
Comparison
of
the
miss
rate
for
the
programs
from
the
SPEC
CPU
2017
benchmark
suite.}
\end{figure}

In
Fig.
\ref{fig:spec:missrate},
the
overall
miss
rate
for
the
SPEC
CPU
2017
benchmark
suite
with
traditional
caches
vs.
\secache{}
is
shown.
Overall,
the
miss
rate
using
\secache{}
is
only
slightly
increased.

\begin{figure}[H]
	\centering
	\includegraphics[width=\linewidth]{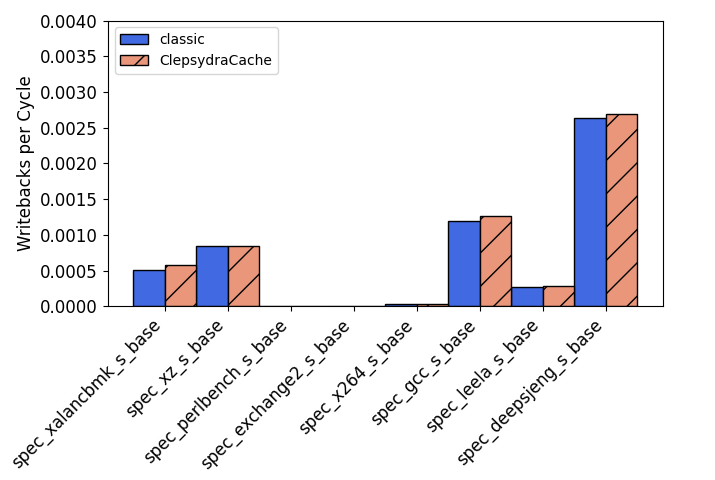}
		\caption{\label{fig:spec:writebacks}
Comparison
of
the
writebacks
per
cycle
for
the
programs
from
the
SPEC
CPU
2017
benchmark
suite.}
\end{figure}

The
number
of
average
writebacks
per
cycle
is
shown
in
Fig.
\ref{fig:spec:writebacks}. \secache{} experiences a
very
small
overhead
in
writebacks
per
cycle
compared
to
a
traditional
cache.

\begin{figure}[H]
	\centering
	\includegraphics[width=\linewidth]{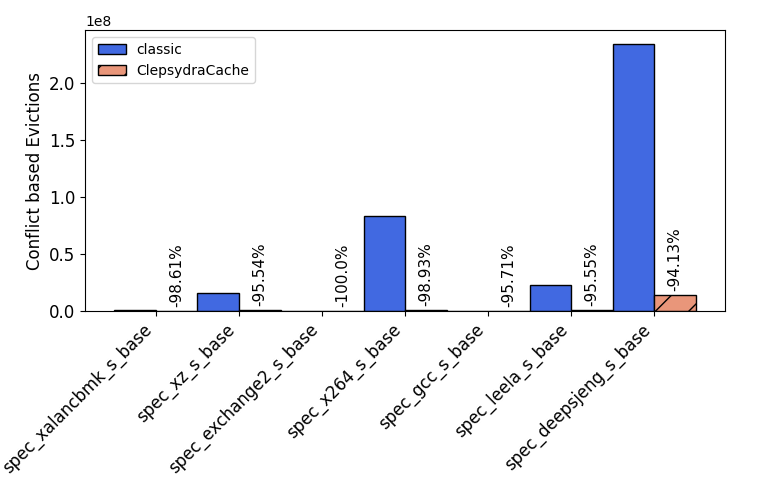}
		\caption{\label{fig:spec:confevict}
Comparison
of
conflict-based
evictions
for
the
programs
from
the
SPEC
CPU
2017
benchmark
suite.}
\end{figure}

Finally,
Fig.
\ref{fig:spec:confevict}
shows
that
\secache{}
reduces
the
number
of
conflict
based
evictions
significantly
for
all
benchmarks
from
the
SPEC
CPU
2017
suite.
The
\textit{exchange2}
benchmark
had
very
few
conflict
based
evictions
to
begin
with
and
\secache{}
managed
to
avoid
all
of
them.
For
the
\textit{gcc}
benchmark,
\secache{}
reduced
the
conflict
based
evictions
from
76,639
to
only
3,290
which
is
a
reduction
of
over
95\%.
The
other
benchmarks
show
similar
behavior.
\section{MiBench Benchmark Results}
\label{app:mibench}

This
Appendix
shows
the
performance
graphs
for
the
MiBench
benchmarks
for
our
Gem5
evaluation.

\begin{figure}[H]
	\centering
	\includegraphics[width=0.75\linewidth]{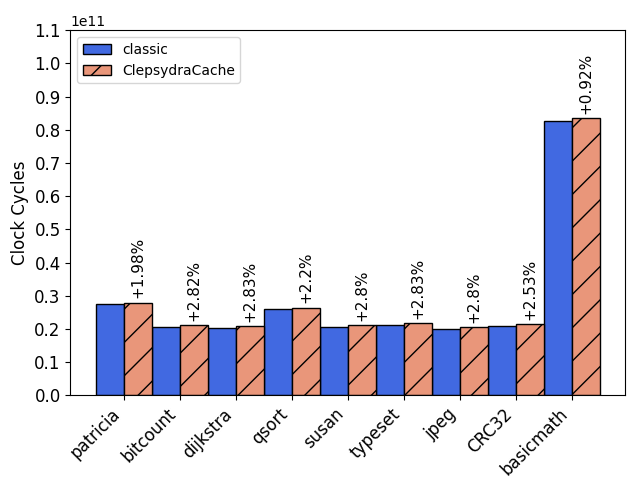}
		\caption{\label{fig:mibench:clock}
Comparison
of
runtime
for
the
programs
from
the
MiBench
benchmark
suite,
including
system
boot
and
shutdown.}
\end{figure}

Fig.
\ref{fig:mibench:clock}
shows
the
runtime
in
clock
cycles
for
the
benchmarks
from
the
MiBench
suite.
\secache{}
encounters
a
performance
overhead
between
0.92\%
and
2.83\%
with
an
average
of
2.4\%.

\begin{figure}[H]
	\centering
	\includegraphics[width=0.75\linewidth]{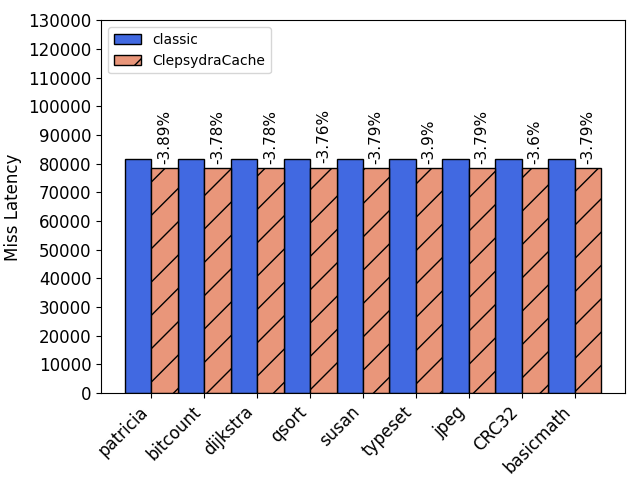}
		\caption{\label{fig:mibench:misslatency}
Comparison
of
the
average
miss
latency
for
the
programs
from
the
MiBench
benchmark
suite,
including
system
boot
and
shutdown.}
\end{figure}

Similar
to
the
other
benchmark
suites
evaluated
in
this
paper,
the
average
miss
latency
reduces
for
all
benchmarks
from
the
MiBench
suite
as
shown
in
Fig.
\ref{fig:mibench:misslatency}
.
The
decrease
is
very
consistent
throughout
the
benchmarks
with
about
3.8\%.

\begin{figure}[H]
	\centering
	\includegraphics[width=0.75\linewidth]{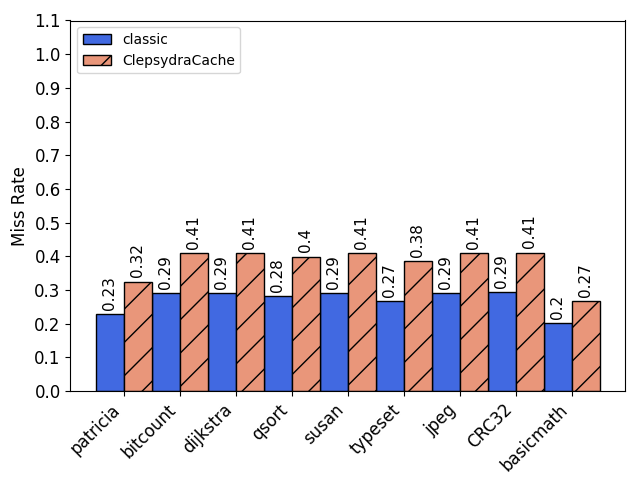}
		\caption{\label{fig:mibench:missrate}
Comparison
of
the
miss
rate
for
the
programs
from
MiBench
benchmark
suite,
including
system
boot
and
shutdown.}
\end{figure}

The
miss
rate
for
the
MiBench
benchmarks
is
shown
in
Fig
\ref{fig:mibench:missrate}. Matching the results from the
other
evaluated
benchmark
suites,
MiBench
encounters
a
slightly
increased
miss
rate
for
all
benchmarks.
The
highest
increase
is
thereby
12\%
and
the
lowest
7\%.

\begin{figure}[H]
	\centering
	\includegraphics[width=0.75\linewidth]{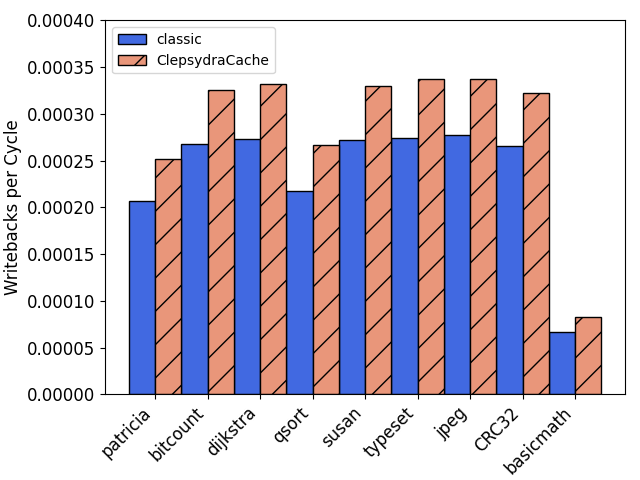}
		\caption{\label{fig:mibench:writebacks}
Comparison
of
the
writebacks
per
cycle
for
the
programs
from
the
MiBench
benchmark
suite,
including
system
boot
and
shutdown.}
\end{figure}

Fig.
\ref{fig:mibench:writebacks}
shows
the
average
number
of
writebacks
per
cycle
for
\secache{}
in
comparison
to
a
traditional
cache
architecture.
The
increase
is
small
for
all
benchmarks.

\begin{figure}[H]
	\centering
	\includegraphics[width=0.75\linewidth]{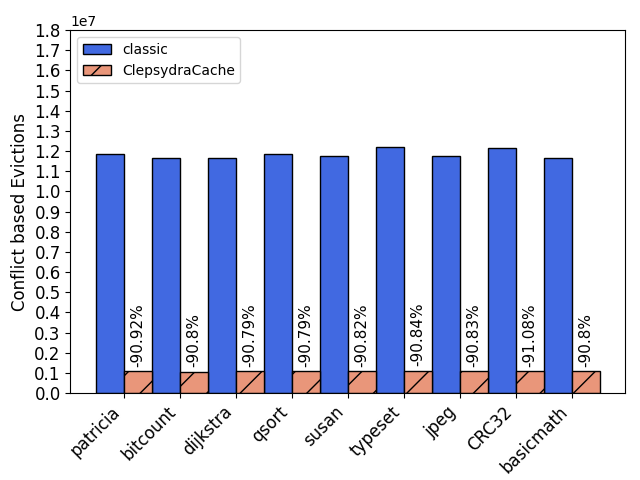}
		\caption{\label{fig:mibench:confevict}
Comparison
of
conflict-based
evictions
for
the
programs
from
the
MiBench
benchmark
suite,
including
system
boot
and
shutdown.}	
\end{figure}

In
Fig.
\ref{fig:mibench:confevict},
the
number
of
conflict
based
evictions
for
the
MiBench
suite
is
depicted.
The
results
are
again
very
consistent
throughout
the
benchmarks,
achieving
a
reduction
of
conflict
based
evictions
by
about
90\%
for
each
benchmark.

\end{document}

%% file: tikz/cache_addressing2.tex

\begin{tikzpicture}
\draw[fill=my_gray_light, rounded corners] (0,0) rectangle (12,-4);

\node[left] at (0, -.625) {\large \textsc{Row 1}};
\node[left] at (0, -1.625) {\large \textsc{Row 2}};
\node[left] at (0, -3.375) {\large \textsc{Row N}};

\node[above] at (1.5, 0) {\large \textsc{Way 1}};
\draw[fill=white, rounded corners] (.25,-.25) rectangle (2.75,-1);
\draw[fill=white, rounded corners] (.25,-1.25) rectangle (2.75,-2);
\node[below] at (1.5, -2) {\textbf{\vdots}};
\draw[fill=white, rounded corners] (.25,-3) rectangle (2.75,-3.75);

\node[above] at (4.5, 0) {\large \textsc{Way 2}};
\draw[fill=white, rounded corners] (3.25,-.25) rectangle (5.75,-1);
\draw[fill=white, rounded corners] (3.25,-1.25) rectangle (5.75,-2);
\node[below] at (4.5, -2) {\textbf{\vdots}};
\draw[fill=white, rounded corners] (3.25,-3) rectangle (5.75,-3.75);

\node[above] at (7.5, 0) {\large \textsc{Way 3}};
\draw[fill=white, rounded corners] (6.25,-.25) rectangle (8.75,-1);
\draw[fill=white, rounded corners] (6.25,-1.25) rectangle (8.75,-2);
\node[below] at (7.5, -2) {\textbf{\vdots}};
\draw[fill=white, rounded corners] (6.25,-3) rectangle (8.75,-3.75);

\node[above] at (10.5, 0) {\large \textsc{Way 4}};
\draw[fill=white, rounded corners] (9.25,-.25) rectangle (11.75,-1);
\draw[fill=white, rounded corners] (9.25,-1.25) rectangle (11.75,-2);
\node[below] at (10.5, -2) {\textbf{\vdots}};
\draw[fill=white, rounded corners] (9.25,-3) rectangle (11.75,-3.75);

%
\draw[draw=none, rounded corners, pattern=north west lines, pattern color=my_blue_light80] (.25,-.25) rectangle (2.75,-1);
\draw[draw=none, rounded corners, pattern=north west lines, pattern color=my_blue_light80] (3.25,-.25) rectangle (5.75,-1);
\draw[draw=none, rounded corners, pattern=north west lines, pattern color=my_blue_light80] (6.25,-.25) rectangle (8.75,-1);
\draw[draw=none, rounded corners, pattern=north west lines, pattern color=my_blue_light80] (9.25,-.25) rectangle (11.75,-1);

\draw[draw=blue, rounded corners] (.125,-.125) rectangle (11.875,-1.125);

\filldraw (1.5,-.625) circle (5pt);
\draw[line width=3pt] (1.5,-.625) -- (1.9,-1.75);
\draw[draw=none, fill=orange!60!black, text=white] (1,-1.75) rectangle (3,-2.5) node[pos=.5] {\Large \textsc{Flags}};
\draw[draw=none, fill=green!40!black, text=white] (3,-1.75) rectangle (5.75,-2.5) node[pos=.5] {\Large \textsc{Tag}};
\draw[draw=none, fill=red!80!black, text=white] (5.75,-1.75) rectangle (9,-2.5) node[pos=.5] {\Large \textsc{Data}};
\draw[line width=3pt, rounded corners] (1,-1.75) rectangle (9,-2.5);

%
\node[left] at (2.5,-4.5) {\Large \textsc{Address:}};
\draw[draw=none,fill=green!40!black, text=white] (2.5,-4.25) rectangle (5.75,-4.75) node[pos=.5] {\Large \textsc{Tag}};
\draw[draw=none,fill=my_blue_light80, text=white] (5.75,-4.25) rectangle (8,-4.75) node[pos=.5] {\Large \textsc{Index}};
\draw[draw=none,fill=gray, text=white] (8,-4.25) rectangle (9.5,-4.75) node[pos=.5] {\Large \textsc{Ofst}};

\end{tikzpicture}

%% file: tikz/skewed_cache.tex

\begin{tikzpicture}
\draw[fill=my_gray_light, rounded corners] (0,0) rectangle (12,-4);

\node[left] at (0, -.625) {\large \textsc{Row 1}};
\node[left] at (0, -1.625) {\large \textsc{Row 2}};
\node[left] at (0, -3.375) {\large \textsc{Row N}};

\node[above] at (1.5, 0) {\large \textsc{Way 1}};
\draw[fill=white, rounded corners] (.25,-.25) rectangle (2.75,-1);
\draw[fill=white, rounded corners] (.25,-1.25) rectangle (2.75,-2);
\node[below] at (1.5, -2) {\textbf{\vdots}};
\draw[fill=white, rounded corners] (.25,-3) rectangle (2.75,-3.75);

\node[above] at (4.5, 0) {\large \textsc{Way 2}};
\draw[fill=white, rounded corners] (3.25,-.25) rectangle (5.75,-1);
\draw[fill=white, rounded corners] (3.25,-1.25) rectangle (5.75,-2);
\node[below] at (4.5, -2) {\textbf{\vdots}};
\draw[fill=white, rounded corners] (3.25,-3) rectangle (5.75,-3.75);

\node[above] at (7.5, 0) {\large \textsc{Way 3}};
\draw[fill=white, rounded corners] (6.25,-.25) rectangle (8.75,-1);
\draw[fill=white, rounded corners] (6.25,-1.25) rectangle (8.75,-2);
\node[below] at (7.5, -2) {\textbf{\vdots}};
\draw[fill=white, rounded corners] (6.25,-3) rectangle (8.75,-3.75);

\node[above] at (10.5, 0) {\large \textsc{Way 4}};
\draw[fill=white, rounded corners] (9.25,-.25) rectangle (11.75,-1);
\draw[fill=white, rounded corners] (9.25,-1.25) rectangle (11.75,-2);
\node[below] at (10.5, -2) {\textbf{\vdots}};
\draw[fill=white, rounded corners] (9.25,-3) rectangle (11.75,-3.75);

\draw[draw=none, rounded corners, pattern=north east lines, pattern color=my_red_light60] (.25,-.25) rectangle (2.75,-1);
\draw[draw=none, rounded corners, pattern=north east lines, pattern color=my_red_light60] (3.25,-1.25) rectangle (5.75,-2);
\draw[draw=none, rounded corners, pattern=north east lines, pattern color=my_red_light60] (6.25,-3) rectangle (8.75,-3.75);
\draw[draw=none, rounded corners, pattern=north east lines, pattern color=my_red_light60] (9.25,-1.25) rectangle (11.75,-2);

\draw[draw=none, rounded corners, pattern=north west lines, pattern color=my_blue_light80] (.25,-1.25) rectangle (2.75,-2);
\draw[draw=none, rounded corners, pattern=north west lines, pattern color=my_blue_light80] (3.25,-1.25) rectangle (5.75,-2);
\draw[draw=none, rounded corners, pattern=north west lines, pattern color=my_blue_light80] (6.25,-.25) rectangle (8.75,-1);
\draw[draw=none, rounded corners, pattern=north west lines, pattern color=my_blue_light80] (9.25,-3) rectangle (11.75,-3.75);

\draw[fill=white, rounded corners, pattern=north west lines, pattern color=my_blue_light80] (2,-4.25) rectangle (2.5,-4.75) node[pos=.5, right=10pt] {\large \textsc{Address 1}};
\draw[fill=white, rounded corners, pattern=north east lines, pattern color=my_red_light60] (7,-4.25) rectangle (7.5,-4.75) node[pos=.5, right=10pt] {\large \textsc{Address 2}};
 
\end{tikzpicture}